\documentclass[journal]{IEEEtran}
\usepackage[cmex10]{amsmath}
\usepackage{eqparbox}
\usepackage[usenames, dvipsnames]{color}
\usepackage{mathtools,amssymb,bm,mathabx}
\usepackage{amstext}
\usepackage{amssymb}
\usepackage{graphicx}
\usepackage{color}
\usepackage{booktabs}
\usepackage{longtable}
\usepackage{multicol}
\usepackage{multirow}
\usepackage{amsfonts}
\usepackage{dsfont}
\usepackage{array}
\usepackage[ruled]{algorithm}
\usepackage{algorithmic}
\usepackage{cases}
\usepackage{pgfplots}
\usepackage{accents}
\usepackage{caption}
\usepackage[footnotesize]{subfigure}
\usepackage{amsthm,xparse}
\DeclareCaptionLabelSeparator{periodspace}{.\quad}
\usepackage{float}
\usepackage{cite}
\usepackage [autostyle, english = american]{csquotes}
\usepackage{hyperref}
\theoremstyle{remark}

\newcommand{\mx}[1]{\mathbf{#1}}
\newcommand{\bs}[1]{\boldsymbol{#1}}

\newcommand\ASTART{\bigskip\noindent\begin{minipage}[b]{0.5\linewidth}}
	
	\newcommand\AENDSKIP{\end{minipage}\bigskip}
\newcommand\AEND{\end{minipage}}
\ifCLASSOPTIONcaptionsoff
\usepackage[nomarkers]{endfloat}

\fi
\theoremstyle{plain}
\newtheorem{thm}{\textbf{Theorem}}
\newtheorem{lem}{\textbf{Lemma}}

\theoremstyle{definition}
\newtheorem{defn}{\textbf{Definition}}

\theoremstyle{remark}
\newtheorem{rem}{\bf Remark}

\newcommand*{\rom}[1]{\expandafter\@slowromancap\romannumeral #1@}
\def\change{black}

\newcommand{\RN}[1]{%
\textup{\uppercase\expandafter{\romannumeral#1}}%
}

\usepackage[labelfont=bf,font=small]{caption}
\renewcommand{\triangleq}{\mathbin{\setstackgap{S}{0pt}\stackMath\Shortstack{\smalltriangleup\\ =}}}
\usepackage[usestackEOL]{stackengine} 
\usepackage{standalone}
\graphicspath{{./Figures/}} 
\begin{document}
%
\title{Exploiting Subspace Prior Information in Dynamic Matrix Completion}
\title{Dynamic Matrix Completion: Harnessing the Power of Subspace Priors}
\title{Matrix Completion Empowered by Arbitrary Subspace Prior Knowledge}
\title{Subspace-Informed Matrix Completion}
\author{
 Hamideh.Sadat~Fazael~Ardakani, Sajad~Daei, Arash Amini , Mikael Skoglund, Gabor Fodor\\
}

\maketitle

\begin{abstract}

In this work, we consider the matrix completion problem, where the objective is to reconstruct a low-rank matrix from a few observed entries. A commonly employed approach  involves nuclear norm minimization. For this method to succeed, the number of observed entries needs to scale at least proportional to both the rank of the ground-truth matrix and the coherence parameter. While the only prior information is oftentimes the low-rank nature of the ground-truth matrix, in various real-world scenarios, additional knowledge about the ground-truth low-rank matrix is available. For instance, in collaborative filtering, Netflix problem, and dynamic channel estimation in wireless communications, we have partial or full knowledge about the signal subspace  in advance.
Specifically, we are aware of some subspaces that 
form multiple angles with the column and row spaces of the ground-truth matrix. Leveraging this valuable information has the potential to significantly reduce the required number of observations. To this end, we introduce a  multi-weight nuclear norm optimization problem that concurrently promotes the low-rank property as well the information about the available subspaces.
The proposed weights are tailored to penalize each angle corresponding to each basis of the prior subspace independently. We further propose an optimal weight selection strategy by minimizing the coherence parameter of the ground-truth matrix, which is equivalent to minimizing the required number of observations. Simulation results validate the advantages of incorporating multiple weights in the completion procedure. Specifically, our proposed multi-weight optimization problem demonstrates a substantial reduction in the required number of observations compared to the state-of-the-art methods.

\end{abstract}
\section{Introduction}\label{proof.Matrix completion }
Matrix completion, a fundamental concept in data science, consists of  reconstructing  a low-rank matrix $\mathbf{X} \in \mathbb{R}^{n\times n}$ with rank $r\ll n$  based on a limited set of noisy observed entries $\mathbf{Y} = \mathcal{R} _\Omega (\mathbf{X} + \mathbf{E}) \in \mathbb{R}^{n \times n } $ \cite{candes2009exact,recht2011simpler}. 
Here, $\mathbf{E}\in\mathbb{R}^{n\times n}$ stands for the noise matrix, $\Omega$ is the index set of observed entries, and the observation operator $\mathcal{R} _\Omega(\cdot)$ for a matrix $\mathbf{Z} \in \mathbb{R}^{n \times n}$ is defined as
\begin{align}
& \mathcal{R} _\Omega (\mx{Z}) \triangleq \sum_{i,l=1}^{n} \tfrac{\epsilon_{il}}{p_{il}}\, \langle \mathbf{ Z }, \mathbf{e}_i\mathbf{e}_l^{T} \rangle \, \mathbf{e}_i\mathbf{e}_{l}^{T}, \label{sampling_opr}
\end{align}	
where $\{\mathbf{e}_i\}_{i=1}^{n}$ are the canonical basis for $\mathbb{R}^n$, 
$p_{il}$ is the probability of observing $(i,l)$-th element of the matrix, and $\epsilon_{il}$ is a Bernoulli random variable taking values $1$ and $0$ with probabilities $p_{il}$ and $1-p_{il}$, respectively. The expected size of the observed set is therefore, $|\Omega|=\sum_{i,l=1}^n p_{il}$. While many matrices 
satisfy the observations, it is well-established that under certain conditions, the matrix with the lowest rank is unique \cite{candes2010matrix}. Hence, typically, to promote the low-rank feature, the following rank minimization is employed:  
\begin{align}
\underset{\mathbf {Z} \in \mathbb{R}^{n \times n}}{\rm min}~ {\rm rank}{(\mathbf{Z})}, \quad{\rm s.t.}~\|\mathbf{Y} - \mathcal{R} _\Omega (\mathbf{Z})\|_F\le e , \label{min rank(z)}
\end{align}	
where $e$ is an upper-bound for $\|\mathcal{R} _\Omega(\mathbf{E})\|_F$.

Since \eqref{min rank(z)} is generally 
an NP-hard problem and intractable, it is commonly replaced with the following surrogate optimization:
\begin{align}
\underset{\mathbf {Z} \in \mathbb{R}^{n \times n}}{\rm min}~ \|\mathbf{Z}\|_*, \quad {\rm s.t.}~\|\mathbf{Y} - \mathcal{R} _\Omega (\mathbf{Z})\|_F\le e, \label{nuclear_prob}
\end{align}	
where $\| \cdot \|_*$ is called the nuclear norm which computes the sum of the singular values of a matrix. It is known that the problem \eqref{nuclear_prob} is a convex relaxation  of \eqref{min rank(z)} \cite{recht2010guaranteed}. 

Traditionally, the only considered prior information in this context is the inherent low-rank nature of the ground-truth matrix. However, in many applications such as quantum state tomography \cite{gross2010quantum}, magnetic resonance imaging\cite{haldar2010spatiotemporal,zhao2010low}, collaborative filtering \cite{srebro2010collaborative}, exploration seismology \cite{aravkin2014fast}, the Netflix problem \cite{bennett2007netflix} and wireless communications \cite{li2017millimeter,shen2015joint,valduga2018transceiver}, there are available prior knowledge about the solution in form of the row and column subspaces of the ground-truth matrix $\mathbf{X}$. For instance, in the Netflix problem, prior evaluations of the movies by the referees can provide prior information about the ground-truth subspaces of the Netflix matrix. Further, in sensor network localization \cite{so2007theory}, some information about the position of the sensors can be exploited as available prior knowledge (c.f. \cite[Section I]{daei2018optimal} and \cite[Section I.A]{eftekhari2018weighted} for more applications). In time-varying frequency division duplexing (FDD) in massive multiple-input multiple-output (MIMO) wireless communications, the velocity information of mobile users can provide subspace prior information which can be exploited to reduce the training uplink overhead and consequently enhance the spectral efficiency (see Section \ref{sec:FDD} for more details).

The aforementioned prior subspace information often appear in the form of $r'$-dimensional column and row  subspaces (denoted by $ \widetilde{\bm{\mathcal{U}}}_{r'} $ and $ \widetilde{\bm{\mathcal{V}}}_{r'} $) forming angles with the column and row $r$-dimensional spaces of the ground-truth matrix $\mathbf{X}$, denoted by $\bm{\mathcal{U}}_r$ and  $\bm{\mathcal{V}}_r$, respectively. To incorporate this prior information into the recovery procedure, we propose the following [tractable] problem for low-rank matrix completion:
\begin{align}
&\underset{\mathbf{Z} \in \mathbb{R}^{n \times n}}{\rm min}~ \|\mathbf {Q}_{\widetilde{\bm{\mathcal{U}}}_{r'}} \mathbf{Z} \mathbf {Q}_{\widetilde{\bm{\mathcal{V}}}_{r'}} \|_*  \nonumber \\
&{\rm s.t.}~\|\mathbf {Y} - \mathcal{R} _\Omega (\mathbf Z)\|_F \le e, \label{prior_nuclear_prob}
\end{align}	
where
\begin{align}
\mathbf {Q}_{\widetilde{\bm{\mathcal{U}}}_{r'}}~\triangleq~\widetilde{\mathbf{U}}_{r'} \bs{\Lambda} \widetilde{\mathbf{U}}_{r'}^H + \mathbf{P}_{\widetilde{\bm{\mathcal{U}}}_{r'}^{\perp}}, \quad
\mathbf {Q}_{\widetilde{\bm{\mathcal{V}}}_{r'}}~\triangleq~\widetilde{\mathbf{V}}_{r'} \bs{\Gamma} \widetilde{\mathbf{V}}_{r'}^H + \mathbf{P}_{\widetilde{\bm{\mathcal{V}}}_{r'}^{\perp}},
\end{align}	
and $\bs{\Lambda}, \bs{\Gamma} \in\mathbb{R}^{r^\prime \times r^\prime}$ are diagonal matrices, the diagonal elements of which belong to the interval $[0,1]$; besides, $\widetilde{\mathbf{U}}_{r'},\widetilde{\mathbf{V}}_{r'} \in \mathbb R^{n\times r'}$ indicate some bases for the subspaces $\widetilde{\bm{\mathcal{U}}}_{r'}$ and $\widetilde{\bm{\mathcal{V}}}_{r'}$, respectively.  Also, the orthogonal projection matrix is defined by $\mathbf{P}_{\widetilde{\bm{\mathcal{U}}}_{r'}^{\perp}}\triangleq\mathbf{I}_{n}-\widetilde{\mathbf{U}}_{r'} \widetilde{\mathbf{U}}_{r'}^H$, where $\mathbf{I}_n$ is the identity matrix of size $n$. In the special cases where $\bs{\Lambda} = \bs{\Gamma} = \mathbf{I}_{r'}$, the problem \eqref{prior_nuclear_prob} simplifies to the standard nuclear norm minimization \eqref{nuclear_prob}. 
The diagonal entries of the matrix $\bs{\Lambda}$ are influenced by the accuracy of existing prior information regarding the column space of matrix $\mx{X}$ and are manifested through the principal angles between $\widetilde{\bm{\mathcal{U}}}_{r^\prime}$ and ${\bm{\mathcal{U}}}_{r}$. Specifically, these angles represent the correlation between orthonormal bases of the subspaces, denoted by $\mx{U}_r\in\mathbb{R}^{n\times r}$ and $\widetilde{\mx{U}}_{r^\prime}\in \mathbb{R}^{n\times r^\prime}$. Correspondingly, analogous assertions apply to the matrix $\boldsymbol{\Gamma}$, reflecting the accuracy of prior knowledge concerning the row space of $\mx{X}$.
As an example, when the principal angle between the estimated and true basis vectors (or directions) increases, the accuracy of that direction estimate diminishes and thus, the assigned weight to that direction estimate should intuitively be large (near $1$). Establishing the optimal theoretical connection between the principal angles and the weights in the optimization problem \eqref{prior_nuclear_prob} poses a significant challenge, and addressing this challenge constitutes the main focus of this paper.

\subsection{Prior Arts}\label{sec.relatedwork}
In this section, we provide an overview of existing approaches that have incorporated additional prior information into matrix completion. Several studies \cite{rao2015collaborative,srebro2010collaborative,zhou2012kernelized} present modified versions of trace-norm regularization to take into account the nonuniform sampling distribution. This is in contrast to the uniform sampling used in the unweighted version presented in \eqref{nuclear_prob}, and the proposed modification is expressed as:
\begin{align}\label{5}
\|\mathbf {X} \|_{\rm tr} \triangleq \|{\rm diag}(\sqrt{\mathbf {p}})\mathbf {X }{\rm diag}(\sqrt{\mathbf {q}}) \|_*,
\end{align}
in which $p(i)$ and $q(l)$ indicate the probabilities that the $i$-th row and the $l$-th column of the matrix are observed, respectively. A limitation of this modeling is its failure to consider additional bases and the angles between the prior and ground-truth subspaces. It solely focuses on the presence probability of specific bases within the ground-truth subspace.

In \cite{angst2011generalized}, \cite{jain2013provable} and \cite{xu2013speedup}, the subspace prior information is in the form of completely knowing some directions (bases of subspace) in the row and column subspaces of $\mathbf{X}$. They assign some heuristic weights to penalize these specific directions based on the given subspace prior information.

In \cite{mohan2010reweighted}, the authors explore the minimization of a re-weighted trace norm through an iterative heuristic method, analyzing its convergence. Such approaches seek to leverage prior information derived from previous estimates and are performed across multiple steps. Notably, this form of prior information constitutes an additional estimate of the ground-truth low-rank matrix, rather than having an alternative subspace distinct from the ground-truth subspace considered in \eqref{prior_nuclear_prob}. 


Aravkin et al. present an iterative algorithm in \cite{aravkin2014fast} aimed at integrating subspace prior information into low-rank matrix completion. They assume that there are column and row subspaces $ \widetilde{\bm{\mathcal{U}}}_{r}$ and $ \widetilde{\bm{\mathcal{V}}}_{r}$ that form $r$ principal angles with the ground-truth subspaces $ {\bm{\mathcal{U}}}_{r}$ and $ {\bm{\mathcal{V}}}_{r}$, respectively.  More precisely, they assign the weights $\lambda$, $\gamma$ to the whole prior subspaces and solve the following optimization: 
\begin{align}\label{6}
\underset{\mathbf {Z} \in \mathbb{R}^{n\times n}}{\rm min} ~ \| \mathbf {Q}_{\widetilde{\bm{\mathcal{U}}}_{r}} \mathbf {Z }\mathbf {Q}_{\widetilde{\bm{\mathcal{V}}}_{r}}\|_* , \quad {\rm s.t.}~\mathbf{ Y} = \mathcal{R}_{\Omega} (\mathbf {Z}), \nonumber
\end{align}
where 
\begin{align}
\mathbf {Q}_{\widetilde{\bm{\mathcal{U}}}_{r}}~\triangleq~ \lambda \mathbf{P}_{\widetilde{\bm{\mathcal{U}}}_{r}} + \mathbf{P}_{\widetilde{\bm{\mathcal{U}}}_{r}^{\perp}}, \quad \mathbf {Q}_{\widetilde{\bm{\mathcal{V}}}_{r}}~\triangleq~ \gamma \mathbf{P}_{\widetilde{\bm{\mathcal{V}}}_{r}} + \mathbf{P}_{\widetilde{\bm{\mathcal{V}}}_{r}^{\perp}}.
\end{align}	
Their approach to determine the weights $\lambda$, $\gamma$ is rather heuristic and the theoretical link between maximum principal angles within the  subspace priors and the ground-truth subspaces is not obtained.

Eftekhari et al. in \cite{eftekhari2018weighted} consider the same optimization problem as in \eqref{6} and prove that the number of required observations can be reduced compared to the standard nuclear norm minimization in the presence of subspace prior information. Their approach assigns a single weight to the whole prior subspaces. The optimal weight is chosen by minimizing the coherence parameter of the ground-truth matrix that is a measure of how correlated or aligned the observed columns and rows of a matrix with the canonical bases are. Since the required number of observations for matrix completion is directly proportional to the coherence parameter, the mentioned strategy leads to minimizing the sampling rate. Their method is limited to situations where the prior subspaces closely align with the ground-truth subspaces, making it unsuitable for leveraging diverse angular ranges between the prior and ground-truth subspaces. Additionally, they establish a relationship between the maximum principal angle and the optimal weight. However, penalizing the entire subspaces with a single weight lacks justification, given that directions within subspaces exhibit different angles relative to those of the ground-truth matrix. Their solution falls short of fully exploiting the prior subspace information, as it focuses solely on the maximum principal angle rather than  all angles.

In \cite{ardakani2022multi}, the authors introduce a weighting strategy, similar to \eqref{prior_nuclear_prob}, for the low-rank matrix recovery problem. Their method involves selecting weights to alleviate the restricted isometry property (RIP) condition of the measurement operator. Unfortunately, this approach is not universally applicable to matrix completion problems, as many measurement operators, including those within the matrix completion framework, do not satisfy the RIP condition. {\color{\change}Furthermore, \cite{ardakani2022multi} does not provide a detailed and explicit application demonstrating the accessibility of subspace prior information.}

In \cite{daei2018optimal}, the optimization problem \eqref{prior_nuclear_prob} is explored, and optimal weights are devised using statistical dimension theory, diverging from other approaches focused on maximizing the RIP constant. However, it is worth noting that statistical dimension theory is not universally applicable to various measurement models, including the matrix completion framework. Other works, such as \cite{fathi2021two,ardakani2019greedy}, also consider prior information, albeit with different approaches, and lack analytical results. 
Despite these efforts, 
it is still not clear to what extent prior knowledge can help.
\subsection{Contributions and key differences with prior works}
Next, we highlight our contributions in the following list:
\begin{enumerate}
    \item \textbf{Subspace-informed optimization:} in \eqref{prior_nuclear_prob}, we propose a comprehensive optimization problem that not only does capture the low-rank nature of the ground-truth matrix but also effectively leverages the available subspace prior information in many real-world applications. Our optimization framework is intricately designed to assign distinct weights to each individual basis within the prior subspace. This approach exploits the full utilization of subspace priors in contrast to \cite{eftekhari2018weighted,aravkin2014fast}, in which all bases (directions) in the prior subspace receive the same penalty weight. More explicitly, unlike the methodologies outlined in \cite{eftekhari2018weighted,aravkin2014fast}, the proposed optimization problem \eqref{prior_nuclear_prob} considers different dimensions for prior subspaces (i.e., $r^\prime$) and ground-truth subspaces (i.e., $r$).
    
    \item \textbf{Optimal weight selection:}
we offer a theoretical approach for determining the optimal weights by minimizing a tight upper-bound on the coherence parameter of the ground-truth matrix. This, in turn, results in the minimization of the required sampling rate for matrix completion. In the special case where there is only a single weight, the proposed weight selection strategy provides a better solution than the single-weight solution \cite{eftekhari2018weighted} as the proposed strategy provides a tighter upper-bound for the coherence parameter of the ground-truth matrix.
   
    \item \textbf{Accessible subspace prior information in wireless communications:}
for dynamic channel estimation in FDD massive MIMO wireless communications, we establish a direct connection between the velocity of mobile users and the principal angles that are required in the proposed strategy of optimal weight selection. 
Applying our optimization framework in \eqref{prior_nuclear_prob} to the prior information about users' velocity
presents a great potential to significantly reduce the uplink training overhead  and enhance spectral efficiency.
    \end{enumerate}

In summary, as mentioned above, several studies have explored the impact of prior subspace information; however, we  observe an analytical framework for harnessing prior subspace information
only in \cite{eftekhari2018weighted}.  
Consequently, our benchmark in this analysis is the work of \cite{eftekhari2018weighted}, which employs a single weight to penalize all principal angles. Our theoretical and numerical results verify that our proposed multi-weight method requires fewer samples for matrix completion compared to the single-weight strategy provided in \cite{eftekhari2018weighted}.

\subsection{Outline} 
The paper is organized as follows: in Section \ref{sec.singleweight}, we review the background about the coherence of a matrix and the principal angles between subspaces that are required for our analysis later. In Section \ref{sec.main result}, we present our main results which amount to proposing a non-uniform weighted nuclear norm minimization. In Section \ref{sec:FDD}, an application of exploiting subspace information in wireless communications is investigated. In Sections \ref{sec.simulation}, some numerical results are provided which verify the superior performance of our method. Finally, in Section \ref{sec.conclusion}, the paper is concluded.

\subsection{Notations}
Throughout the paper, scalars are indicated by lowercase letters, vectors by lowercase boldface letters, and matrices by uppercase letters. The trace and Hermitian of a matrix are shown as ${\rm Tr}(\cdot)$ and $(\cdot)^{\rm H}$, respectively. The Frobenius inner-product is defined as $\langle\mathbf {A} , \mathbf {B}\rangle_F = {\rm Tr}(\mathbf {A} \mathbf{B}^{H})$. $\| \cdot \|$ denotes the spectral norm and $\mathbf {X} \succcurlyeq 0$ means that $\mathbf {X}$ is a semidefinite matrix. We describe the linear operator $\mathcal{A}:\mathbb{R}^{m \times n}\longrightarrow \mathbb{R}^p$ as
$\mathcal{A} (\mx{X}) = [\langle \mathbf {X} , \mathbf{A}_1 \rangle_F , \cdots ,\langle \mathbf {X }, \mathbf{A}_p \rangle_F]^T$
where $\mathbf{A}_i\in \mathbb{R}^{m \times n}$. The adjoint operator of $\mathcal{A}$ is defined as $\mathcal{A}^*(\mx{y}) = \sum_{i=1}^{p} y_i\mathbf{A}_i$ and $\mathcal{I}$ is the identity linear operator, i.e. 
$\mathcal{I}\mathbf {X}=\mathbf{X}$.
$\rm j$ is the complex imaginary unit. $J_0(\cdot)$ is the Bessel function of zero kind and is defined as $J_0(x)\triangleq \tfrac{1}{2\pi}\int_{-\pi}^{\pi} {\rm e}^{-{\rm j} x \sin(t)}{\rm d}t$. The operation ${\rm diag}(\mathbf{x})$ transforms a vector $\mathbf{x} \in \mathbb{R}^{n\times 1}$ into a diagonal matrix of size $n\times n$ with the elements of $\mathbf{x}$ as its diagonal entries, and all other elements set to zero. The $i$-th element of a vector $\mathbf{x} \in \mathbb{R}^{n\times 1}$ is represented by $x(i)$. Additionally, the element in the $i$-th row and $l$-th column of a matrix $\mathbf{A}$ is denoted by $A_{i,l}$. $\mathbf{e}_i\in\mathbb{R}^{n}$ is the canonical vector having $1$ in the $i$-th location and zero elsewhere. $a\vee b$ means $\max\{a,b\}$ for two arbitrary scalars $a$ and $b$. $ A \gtrsim B$ means that $A$ is greater than $B$ up to a constant factor.

\section{Background}\label{sec.singleweight}
This section is dedicated to explaining concepts related to subspace information and the angles between two subspaces. Additionally, we present the strategy of single-weight penalization as outlined in \cite{eftekhari2018weighted}.

We first provide a definition for the coherence parameter of a subspace in $\mathbb{R}^n$ which serves as a metric for our weight selection method later.
\begin{defn}[\cite{daei2018optimal}] Assume $\bm{\mathcal{U}}_r \subseteq \mathbb{R}^n$ is an $r$-dimensional subspace and $\mathbf{P}_{\bm{\mathcal{U}}_r}$ indicates the orthogonal projection matrix onto that subspace. Then, the $i$-th coherence parameter of $\bm{\mathcal{U}}_r$ is defined as:
	\begin{align}\label{coherence}
	\mu_i(\bm{\mathcal{U}}_r)\triangleq\tfrac{n}{r}~\| \mathbf{P}_{\bm{\mathcal{U}}_r} \mathbf{e}_i \|^2, i=1,..., n.
	\end{align}
	
 Let ${\mx{U}_r}\in\mathbb{R}^{n\times r}$ and ${\mx{V}_r}\in\mathbb{R}^{n\times r}$ be some bases for the column and row spaces $\bm{\mathcal{U}}_r$ and $\bm{\mathcal{V}}_r$, respectively, and $\mx{X}_r={\mx{U}_r}\bs{\Sigma}_r{\mx{V}_r}^H\in\mathbb{R}^{n\times n}$ be the singular value decomposition (SVD) of the matrix $\mx{X}_r$ with rank $r\ll n$. Let $\mu_i(\bm{\mathcal{U}}_r)$ and $\mu_l(\bm{\mathcal {V}}_r)$ be the $i$-th and $l$-th coherence parameters of the column and row spaces of a matrix $\mx{X}_r\in\mathbb{R}^{n\times n}$ of dimension $r$. Then, the coherence parameter of the matrix $\mx{X}_r$ is defined as
\begin{align}\label{eq:coherence_parameter_of_matrix}
    \eta(\mathbf{X}_r)=\eta(\mathbf{U}_r \mathbf{V}_r^H)=~\max_i\mu_i(\bm{\mathcal{U}}_r) \vee \max_l \nu_l(\bm{\mathcal {V}}_r). 
\end{align}
	To establish the principal angles between subspaces $\bm{\mathcal {U}}$ and $\bm{\mathcal{\widetilde {U}}}$, let $r$ and $r^\prime$ denote the dimensions of $\bm{\mathcal {U}}$ and $\bm{\mathcal{\widetilde {U}}}$, respectively, with $r \leq r'$. There exists $r$ non-increasing principal angles $\bs{\theta}_u\triangleq [\theta_u(1),...,\theta_u(r)]^T$ in the angular range $[0^\circ,90^\circ]^r$ given by:
 \begin{align}
	&\theta_u(r) = {\rm min} \bigg\{{\rm cos}^{-1}(\tfrac{|\langle\mx{u} , \widetilde{\mx{u}}\rangle|}{\| \mathbf {u} \|_2 \| \widetilde{\mathbf{u}} \|_2}) : \mathbf {u} \in \bm{\mathcal{U}} , \widetilde{\mathbf{u}} \in \widetilde{\bm{\mathcal{U}}}\bigg\}=\left\langle \mx{u}_r, \widetilde{\mx{u}}_r  \right\rangle\nonumber\\
 &\theta_u(i) = {\rm min} \bigg\{{\rm cos}^{-1}(\tfrac{|\langle\mx{u} , \widetilde{\mx{u}}\rangle|}{\| \mathbf {u} \|_2 \| \widetilde{\mathbf{u}} \|_2}) : \mathbf {u} \in \bm{\mathcal{U}} , \widetilde{\mathbf{u}} \in \widetilde{\bm{\mathcal{ U}}} \nonumber \\
	&~~~~~\mathbf{u} \perp \mathbf{u}_l ~,~ \widetilde{\mathbf{u}} \perp \widetilde{\mathbf{u}}_l,\, \forall l \in \{i+1, \cdots , r\} \bigg\}, i=1,..., r-1,
	\end{align}
	where $\mathbf{u}_l$ and $\widetilde{\mathbf{u}}_l$ are called the $l$-th principal vectors leading to the $l$-th principal angle $\theta_u(l)$.

\end{defn}

\section{Main results}\label{sec.main result}

In this section, we propose a performance guarantee for the proposed optimization problem \eqref{prior_nuclear_prob} which provides a non-uniform weighting strategy for matrix completion with prior subspace information. Consider $\widetilde{\bm{\mathcal{U}}}_{r'}$ and $\widetilde{\bm{\mathcal{V}}}_{r'}$ as prior subspace information forming angles with $\bm{\mathcal{U}}_{r}$ and $\bm{\mathcal{V}}_{r}$, respectively. We optimize the weights according to the values of the principal angles.
Theorem \eqref{thm2} below provides performance guarantees for both noiseless and noisy matrix completion scenarios, with the uniform weighting strategy emerging as a special case. Note that the rank-$r$ truncated SVD of a matrix $\mathbf{X}\in\mathbb{R}^{n\times n}$ is denoted as $\mathbf{X}_r$. Additionally, the residual matrix is represented by $\mathbf{X}_{r^+} = \mathbf{X} - \mathbf{X}_r$.
\begin{thm} \label{thm2}
For a matrix $\mx{X}\in\mathbb{R}^{n\times n}$, let $\mx{X}_r=\mx{U}_r\bs{\Sigma}_r\mx{V}_r^H$ be the rank r truncated SVD form of $\mx{X}$. Let ${\bm{\mathcal{U}}}_r = {\rm span}(\mathbf{X}_r)={\rm span}(\mx{U}_r)$ and ${\bm{\mathcal{V}}}_r = {\rm span}(\mathbf{X}_r^H)={\rm span}(\mx{V}_r)$ denote the column and row subspaces of $\mathbf{X}_r$ respectively, while the $r'$-dimensional subspaces $\widetilde{\bm{\mathcal {U}}}_{r'}$ and $\widetilde{\bm{\mathcal {V}}}_{r'}$ represent the prior subspaces with the assumption that $r'\ge r$. Suppose the principal angles, arranged in non-increasing order, between these two sets of subspaces are given by:
\begin{align}
\bs{\theta}_u=\angle [\bm{\mathcal{U}}_r,\widetilde{\bm{\mathcal{U}}}_r]\in\mathbb{R}^{r\times 1},~\bs{\theta}_v=\angle [\bm{\mathcal{V}}_r,\widetilde{\bm{\mathcal{V}}}_r]\in\mathbb{R}^{r\times 1}.
\end{align}
 Additionally, let $\breve{\mathbf {U}}$ and $\breve{\mathbf V}$ represent orthonormal bases for ${\rm span}([\mathbf{U}_r , \widetilde{\mathbf{U}}_{r'}])$ and ${\rm span}([\mathbf V_r , \widetilde{\mathbf V}_{r'}])$, respectively. For $\lambda_{i,i}, \gamma_{i,i} \in (0,1]$, if $\hat{\mathbf{X}}$ is a solution to \eqref{prior_nuclear_prob}, then we have:
	\begin{align}
	&\| \hat{\mathbf{ X}} - \mathbf{ X }\|_F \lesssim \tfrac{\| \mathbf{X}_{r^+} \|_*}{\sqrt{p}}+e\sqrt{pn} \label{theory_main} 	 
	\end{align}	
	provided that
	\begin{align}\label{eq.prob_observed}
	&1 \geq p \gtrsim {\rm max}[{\rm log}(\alpha_1 \cdot n),1]\tfrac{\eta(\mathbf{X}_r)r{\rm log}n}{n} \nonumber\\
	&	 \quad \quad \quad \quad ~~~~~~~~~~~~~~{\rm max}\big[\alpha_2^2\big(1+\tfrac{\eta(\breve{\mathbf{U}} \breve{\mathbf{ V}}^H)}
 {\eta(\mathbf{U}_r {\mathbf{V}_r}^H)}\big),1\big]\nonumber\\
	&\alpha_3\leq\tfrac{1}{4}
	\end{align}	
	where $\eta(\breve{\mathbf{U}} \breve{\mathbf{V}}^H)$ is the coherence of $\breve{\mathbf {U}} \breve{\mathbf {V}}^H$ and
\begin{align}
&\alpha_1\triangleq\alpha_1(\theta_u(i),\theta_v(i),\lambda_1(i),\gamma_1(i))=\nonumber \\
& \sqrt{\underset{i}{\rm max}\bigg\{ \tfrac{f^2_1(\lambda_1(i) , \theta_u(i))} {f^2_2(\lambda_1(i) , \theta_u(i))} \bigg\}  \underset{i}{\rm max}\bigg\{\tfrac{f^2_1(\gamma_1(i) , \theta_v(i))} {f^2_2(\gamma_1(i) , \theta_v(i))}\bigg\}}  \nonumber \\
& \alpha_2 \triangleq \alpha_2(\theta_u(i),\theta_v(i),\lambda_1(i),\gamma_1(i)) 
=\nonumber \\
& \sqrt{\underset{i}{\rm max} \Big\{f^2_2(\lambda_1(i) , \theta_u(i))\Big\}  \underset{i}{\rm max} \bigg\{\tfrac{f^2_1(\gamma_1(i) , \theta_v(i))} {f^2_2(\gamma_1(i) , \theta_v(i))}\bigg\}}   + \nonumber \\
& \sqrt{\underset{i}{\rm max}\Big\{f^2_2(\gamma_1(i) , \theta_v(i))\Big\} \underset{i}{\rm max} \bigg\{\tfrac{f^2_1(\lambda_1(i) , \theta_u(i))} {f^2_2(\lambda_1(i) , \theta_u(i))}\bigg\}} \nonumber \\ 
& \alpha_3 \triangleq \alpha_3 (\theta_u(i),\theta_v(i),\lambda_1(i),\lambda_2(i),\gamma_1(i),\gamma_2(i)) 
=\nonumber \\
& \sqrt{\underset{i}{\rm max} \bigg\{\tfrac{f^2_2(1-\lambda^2_{1}(i) , \theta_u(i))}{f^2_2(\lambda_1(i) , \theta_u(i))} \bigg\}} \cdot
\sqrt{ \underset{i}{\rm max} \bigg\{\tfrac{f^2_2(1-\gamma^2_{1}(i) , \theta_v(i))}{f^2_2(\gamma_1(i) , \theta_v(i))} \bigg\} }  \nonumber \\
&- {\rm max}\bigg\{\underset{i}{\rm max}\{\lambda_2(i)-1\},\underset{i}{\rm max}\bigg\{ \tfrac{\lambda_1(i)}{f_2(\lambda_1(i) , \theta_u(i))}-1 \bigg\}\bigg\}\nonumber\\
& -  {\rm max}\bigg\{\underset{i}{\rm max}\{\gamma_2(i)-1\},\underset{i}{\rm max}\bigg\{ \tfrac{\gamma_1(i)}{f_2(\gamma_1(i) , \theta_v(i))}-1 \bigg\}\bigg\},
\end{align}
\end{thm}
in which 
\begin{align}\label{def_fun}
	&f_1(w , \theta) \triangleq\sqrt{w^4 {\rm cos}^2 (\theta) + {\rm sin}^2 (\theta)} \nonumber \\ 
	&f_2(w , \theta) \triangleq\sqrt{w^2 {\rm cos}^2 (\theta) + {\rm sin}^2 (\theta)} ,
	 \end{align} 
$$ \bs{\Lambda} \triangleq \begin{bmatrix}
\bs{\Lambda}_1 = {\rm diag}(\bs{\lambda}_1) \in \mathbb{R}^{r \times r} & \\
& \bs{\Lambda}_2 ={\rm diag}(\bs{\lambda}_2)\in \mathbb{R}^{r^{\prime}-r \times r^{\prime}-r}  \end{bmatrix} $$ and 
$$ \bs{\Gamma} \triangleq \begin{bmatrix}
\bs{\Gamma}_1 =  {\rm diag}(\bs{\gamma}_1) \in \mathbb{R}^{r \times r}  & \\
& \bs{\Gamma}_2 =  {\rm diag}(\bs{\gamma}_2) \in \mathbb{R}^{r^{\prime}-r \times r^{\prime}-r} 
\end{bmatrix} $$
Proof. See Appendix \ref{proof_th}.
\begin{rem}
	If $\bs{\Lambda}=\bs{\Gamma}=\mathbf{I}_{r'}$ then $\mathbf{Q}_{\bm{\widetilde{\mathcal{U}}_{r'}}} = \mathbf{Q}_{\bm{\widetilde{\mathcal{V}}_{r'}}} = \mathbf{I}_{n}$ and the problem reduces to the standard matrix completion \eqref{prior_nuclear_prob}. Also, considering $\bs{\Gamma}=\gamma \mathbf{I}_r$ and $\bs{\Lambda}=\lambda \mathbf{I}_r$, \eqref{prior_nuclear_prob} reduces to the single weighted problem studied in \cite{eftekhari2018weighted}.
\end{rem}
\begin{rem}(Finding optimal weights)
	Our goal is to reduce the number of required observations or alternatively the recovery error in matrix completion problem. Therefore, we choose the optimal weights that minimizes the lower-bound on $p$ in \eqref{eq.prob_observed} and consequently the recovery error in \eqref{theory_main}.
\end{rem}
\begin{rem}
	 In our proposed model, all principal angels between subspaces are assumed to be accessible. Accordingly, this provides higher degrees of freedom to reduce the number of required observations which is in turn enhances the performance accuracy.
\end{rem}

The following section outlines a specific scenario where the entire prior subspace is penalized with a single weight (also called uniform weighting) and indeed the entire prior subspace is treated uniformly.
\subsection{Uniform Weighting}\label{sec:uniform_weighting}
By employing the single-weight strategy provided in \cite{eftekhari2018weighted}, the prior subspaces $\widetilde{\bm{\mathcal{{U}}}}$ and $\widetilde{\bm{\mathcal{{V}}}}$ are of dimensions $r^\prime=r$ and the weighing matrices in \eqref{prior_nuclear_prob} are set as $\bs{\Lambda}=\lambda\mx{I}_{r}\in \mathbb{R}^{r\times r}$ and $\bs{\Gamma}=\gamma \mx{I}_{r}\in\mathbb{R}^{r\times r}$. 
\begin{thm}[\cite{eftekhari2018weighted}]
	 Let $\mathbf{X}_r=\mathbf{U}_r\bs{\Sigma}_r\mathbf{V}_{r}^{H} \in \mathbb R ^{n \times n}$ be the truncated SVD version of matrix $\mathbf {X} \in \mathbb{ R} ^{n\times n}$, for an integer $r \leq n$ and let $\mathbf{ X} _{r^+}=\mathbf {X} - \mathbf{X}_r$ be the residual matrix. Consider $\widetilde{\bm{\mathcal {U}}}_r$ and $\widetilde{\bm{\mathcal {V}}}_r$ as prior subspace information that form $r$ principal angles with the column and row spaces of matrix $\mx{X}_r$, respectively. Additionally, let $\breve{\mathbf {U}}$ and $\breve{\mathbf {V}}$ represent orthonormal bases for ${\rm span}([\mathbf{U}_r , \widetilde{\mathbf{U}}_r])$ and ${\rm span}([\mathbf {V}_r , \widetilde{\mathbf {V}}_r])$, respectively. For $\lambda,\gamma \in (0,1]$, if $\widehat{\mathbf{X}}$ is a solution to \eqref{prior_nuclear_prob}, then 
	 \begin{align}
	 &\| \hat{\mathbf {X}} - \mathbf {X} \|_F \lesssim \tfrac{\| \mathbf{X}_{r^+} \|_*}{\sqrt{p}}+e\sqrt{pn} \label{theory_armin} 	 
	 \end{align}	
	 provided that
	 \begin{align}
	 &1 \geq p \gtrsim {\rm max} [{\rm log}(\alpha_4 \cdot n ),1]. \tfrac{\eta(\mathbf{X}_r) r {\rm log}n}{n}\nonumber\\
	 & \quad ~~~~~~~~~~~{\rm max}[\alpha_5^2(1+\tfrac{\eta(\breve{\mathbf {U}} \breve{\mathbf {V}}^H)}{\eta(\mathbf {U} {\mathbf {V}}^H)}),1] \nonumber 	 \\
	 &\alpha_6 \leq \tfrac{1}{8}  \label{condition_armin}
	 \end{align}	
	 where $\eta(\breve{\mathbf {U}} \breve{\mathbf {V}}^H)$ represents the coherence of $\breve{\mathbf {U}} \breve{\mathbf {V}}^H$ and
	 \begin{align}
 &\alpha_4 
\triangleq
 \tfrac{f_1(\gamma, \theta_{u}(1))f_1(\lambda, \theta_{v}(1))}{f_2(\lambda, \theta_{u}(1))f_2(\gamma, \theta_{v}(1)) }\nonumber\\
&\alpha_5 
\triangleq
\bigg(  \tfrac{f_2(\lambda, \theta_{u}(1))}{f_2(\gamma, \theta_{v}(1))}  + 
 \tfrac{f_2(\gamma, \theta_{v}(1))}{f_2(\lambda, \theta_{u}(1))}  \bigg) \cdot \nonumber\\
&~~~~~~~~~~~~~~~~~~~~\Big( f_1(\lambda, \theta_{u}(1)) + f_1(\gamma, \theta_{v}(1)) \Big) \nonumber \\
&\alpha_6 
\triangleq 
\tfrac{3}{2}\Bigg(\tfrac{\sqrt{1-\lambda^2}~{\rm sin}\theta_u(1)}{f_2(\lambda, \theta_{u}(1))} + 
\tfrac{\sqrt{1-\gamma^2}~{\rm sin}\theta_v(1)}{f_2(\gamma, \theta_{v}(1))}\Bigg).
	\end{align}
	
\end{thm}
\begin{rem}(Comparison with the special case of Theorem \ref{thm2})
In the special case of Theorem \ref{thm2} where all weights are the same, we have
\begin{align}
    \alpha_2= \tfrac{f_2(\lambda, \theta_{u}(1))}{f_2(\gamma, \theta_{v}(1))} \cdot f_1(\gamma, \theta_{v}(1)) + 
 \tfrac{f_2(\gamma, \theta_{v}(1))}{f_2(\lambda, \theta_{u}(1))}   \cdot  f_1(\lambda, \theta_{u}(1)).
\end{align}
By comparing the sample complexity result \eqref{eq.prob_observed} with \eqref{condition_armin}, we observe that $\alpha_5>\alpha_2$. Since the optimal weights are obtained by minimizing the sample complexity, this in turn shows that even in the special case of uniform weighting, our proposed optimal weights leads to lower number of required observations and consequently less recovery errors. 
\end{rem}
\begin{rem}
Setting $\lambda = \gamma = 1$ in problem \eqref{prior_nuclear_prob} simplifies it to the standard unweighted nuclear norm minimization problem \eqref{nuclear_prob}, yielding $\alpha_4 = 1$, $\alpha_5 = 4$, and $\alpha_6 = 0$. Consequently, \eqref{condition_armin} transforms to
\begin{align}
1 \geq p \gtrsim \tfrac{\eta(\mathbf{X}_r)r{\rm log}^2 n }{n}\big( 1+\sqrt{\tfrac{\eta(\breve{\mathbf U} \breve{\mathbf V}^H)}{\eta(\breve{\mathbf {U}}_r \breve{\mathbf V}_r^H)}} \big)
\end{align}
Furthermore, due to the term $\sqrt{\tfrac{\eta(\breve{\mathbf {U}} \breve{\mathbf {V}}^H)}{\eta(\breve{\mathbf {U}}_r \breve{\mathbf {V}}_r^H)}}$, the probability of observing an element in this specific scenario appears to be a bit looser than the sample complexity of conventional matrix completion which is  $1 \geq p \gtrsim \tfrac{\eta (\mathbf{X}_r)r\log^2 n}{n}$, as discussed in \cite{candes2010matrix}.
	\end{rem}
\section{Application in FDD Massive MIMO communications}\label{sec:FDD}

In this section, we present a specific application in wireless communications where subspace prior information is accessible (see \cite{li2017millimeter,shen2015joint,valduga2018transceiver} for more details). We focus on a frequency division duplexing (FDD) multiple-input multiple-output (MIMO) system where a base station (BS) transmits a pilot signal to $K$ users in the downlink. The users then feedback under-sampled versions of their own pilot observations to the BS in the uplink. {\color{\change} If the pilot symbols follow an independent Gaussian distribution, the task of channel estimation is often simplified to a matrix recovery problem, as explored in \cite{ardakani2022multi}. In such cases, one can maximize the RIP constant to optimally exploit the subspace prior information. However, if the pilots are in binary format (consisting of $0$ and $1$), the task of channel estimation in the presence of subspace prior information can not be performed by the approach used in \cite{ardakani2022multi} and transforms into a subspace-informed matrix completion problem. It is also important to note that the theoretical analysis to find the optimal relationship between the velocity of mobile users and the prior subspace is not provided in \cite{ardakani2022multi}. In FDD massive MIMO systems, binary pilots are often preferred over ordinary pilots due to their simplicity and efficiency. Binary pilots, composed of $0$s and $1$s, are easier to generate, transmit, and process, reducing computational complexity and power consumption. Their clear decision boundaries enhance robustness to noise and interference, ensuring more reliable detection and channel estimation. Additionally, binary pilots facilitate efficient feedback mechanisms, enabling compact and rapid transmission of channel state information from users to the BS. This efficiency is crucial in massive MIMO systems where the number of antennas and users is large, making binary pilots a practical choice for maintaining high performance and scalability.} 

\begin{figure}
    \centering
    \includegraphics[scale=.2]{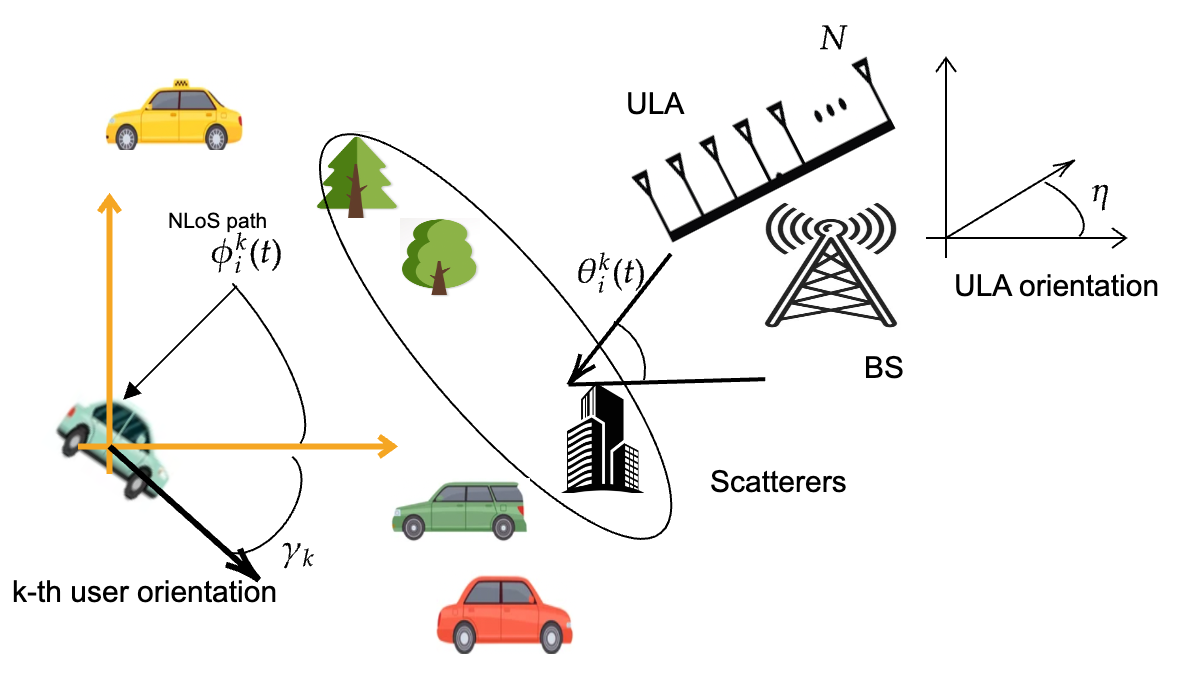}
    \caption{A typical example of exploiting subspace prior information in FDD massive MIMO systems. The BS is equipped with ULA array with $N$ antennas. $\theta_i^k$ and $\phi_i^k$  show the AoD and AoA corresponding to $i$-th path of the channel of $k$-th user. The Doppler frequencies of users provides the subspace prior information required to estimate the channel at the current time.}
    \label{fig:practical_example}
\end{figure}

Mathematically, the resulting observations of all users at the BS can be expressed as
\begin{align}
    \mx{Y}=\mathcal{R}_{\Omega}\Big(\mx{H}(t)+\mx{E}(t)\Big)~\in\mathbb{C}^{N\times K},
\end{align}
where $\Omega$ shows the observed entries with size $|\Omega|=T$, and $\mx{H}(t)\triangleq [\mx{h}_1(t),\ldots \mx{h}_K(t)]$ denotes the downlink channel between the BS and of users at time $t$. User $k$ moves in a direction denoted by $\gamma_k$ and the BS is equipped with uniform linear array (ULA) with $N$ antennas with antenna spacing $d$. The BS is aligned in the direction $\eta$ as shown in Figure \ref{fig:practical_example}. The Doppler frequency corresponding to $k$-th user and $i$-th path is shown by $f_i^k=\tfrac{\nu_k\cos(\phi^k_i)}{\lambda}$ where $\lambda$ denotes carrier wavelength and $\phi_i^k$ is angle of arrival (AoA) corresponding to path $i$ and user $k$. $\nu_k$ is the velocity of $k$-th user. The normalized channel of the $k$-th user at time $t$ is expressed as:
\begin{align}
    \mx{h}_k(t)=\tfrac{1}{\sqrt{N s(t)}}\sum_{i=1}^{s(t)}{\rm e}^{{\rm j}(2\pi f_i^k t+\beta_i^k)} \mx{a}(\theta^k_i)
\end{align}
where $\beta_i$ is the phase of the $i$-th channel path of $k$-th user,  $\theta^k_i$ is the angle of departure (AoD) corresponding to path $i$ and user $k$, $\mx{a}(\theta)\triangleq [1, {\rm e}^{{\rm j}\tfrac{2 \pi d \cos(\eta-\theta)}{\lambda}},..., {\rm e}^{{\rm j}\tfrac{2 \pi d \cos(\eta-\theta)(N-1)}{\lambda}} ] ^{\rm T}\in\mathbb{C}^{N\times 1}$ is the array response vector and $s(t)$ is the total number of scatterers between user $k$ and the BS.
The cross-correlation between the normalized channel of user $k$ at time $t_1$ and user $l$ at time $t_2<t_1$ are obtained by
\begin{align}
\mx{P}(t_1,t_2)\triangleq\mathds{E}\left[\mx{h}_k(t_1)\mx{h}_l^{\rm H}(t_2)\right]  \in\mathbb{C}^{N\times N},
\end{align}
where the above expectation is taken over the random phases and angles. Based on the results provided in \cite{daei2024towards}, the $(p,q)$ element of $\mx{P}(t_1,t_2)$ denoted by ${P}_{(p,q)}(t_1,t_2)$ can be obtained
as follows:
\begin{align}\label{eq:correlation_formula}
    &{P}_{(p,q)}(t_1,t_2)=\nonumber\\
    &\left\{\begin{array}{cc}
         \alpha^\prime J_0(\tfrac{2\pi}{\lambda}\nu_k t_1) J_0(\tfrac{2\pi}{\lambda}\nu_l t_2) J_0(\tfrac{2\pi}{\lambda}d p) J_0(\tfrac{2\pi}{\lambda}d q)& k\neq l \\
          \alpha^\prime J_0(\tfrac{2\pi}{\lambda}\nu_k (t_1-t_2)) J_0(\tfrac{2\pi}{\lambda}d (p-q))& k=l
    \end{array}\right\}
\end{align}
where $\alpha^\prime\triangleq\tfrac{\min(s(t_1), s(t_2))}{\sqrt{s(t_1)s(t_2)}}$ and AoA and AoDs and phases are assumed to be i.i.d. and uniformly distributed between $[-\pi, \pi]$. Now, suppose that $\mx{H}\triangleq [\mx{h}_1(t_1),\ldots, \mx{h}_K(t_1)]$ and $\widetilde{\mx{H}}\triangleq [\mx{h}_1(t_2),\ldots, \mx{h}_K(t_2)]$ are the channel matrices of users at times $t_1$ (current time) and $t_2$ (Previous time), respectively. To obtain the principal angles between the column spaces of $\mx{H}$, i.e., ${\rm span}(\mx{H})$, and the column spaces of $\widetilde{\mx{H}}$ ${\rm span}(\widetilde{\mx{H}})$, we construct $\mathds{E}\mx{H}^H \widetilde{\mx{H}}$ as follows:
\begin{align}\label{eq:H^HHtild}
   \mathds{E}[\mx{H}^H \widetilde{\mx{H}}]= \begin{bmatrix}
        \mathds{E}[\mx{h}_1(t_1)^H \widetilde{\mx{h}}_1(t_2)]&\ldots& \mathds{E}[{\mx{h}}_1(t_1)^H \widetilde{\mx{h}}_K(t_2)]\\
        \vdots&\ddots&\vdots\\
        \mathds{E}[\mx{h}_K(t_1)^H \widetilde{\mx{h}}_1(t_1)]&\ldots&\mathds{E}[\mx{h}_K(t_1)^H \widetilde{\mx{h}}_K(t_2)]
    \end{bmatrix}
\end{align}
By replacing \eqref{eq:correlation_formula} into \eqref{eq:H^HHtild} and taking SVD decomposition, we can have access to the singular values of $ \mathds{E}[\mx{H}^H \widetilde{\mx{H}}]$ which are the cosine of the vector of principal angles between ${\rm span}(\mx{H})$ and ${\rm span}(\widetilde{\mx{H}})$. Similarly, to obtain the principal angles between row spaces of $\mx{H}$ and $\widetilde{\mx{H}}$, we construct $\mathds{E}[\mx{H}\widetilde{\mx{H}}^H]$ as follows:
\begin{align}\label{eq:HHtild^H}
    \mathds{E}[\mx{H} \widetilde{\mx{H}}^H]=\sum_{i=1}^K \mathds{E}[\mx{h}_i(t_1) \mx{h}_i(t_2)^H]
\end{align}
By replacing \eqref{eq:correlation_formula} into \eqref{eq:HHtild^H}, and taking SVD, provides the singular values of $ \mathds{E}[\mx{H} \widetilde{\mx{H}}^H]$ which is equal to the cosine of the vector of principal angles between row spaces of $\mx{H}$ and $\widetilde{\mx{H}}$. 
\section{Simulation Results}\label{sec.simulation}
\begin{figure*}
	\includegraphics[width=3.5in,height=1.5in]{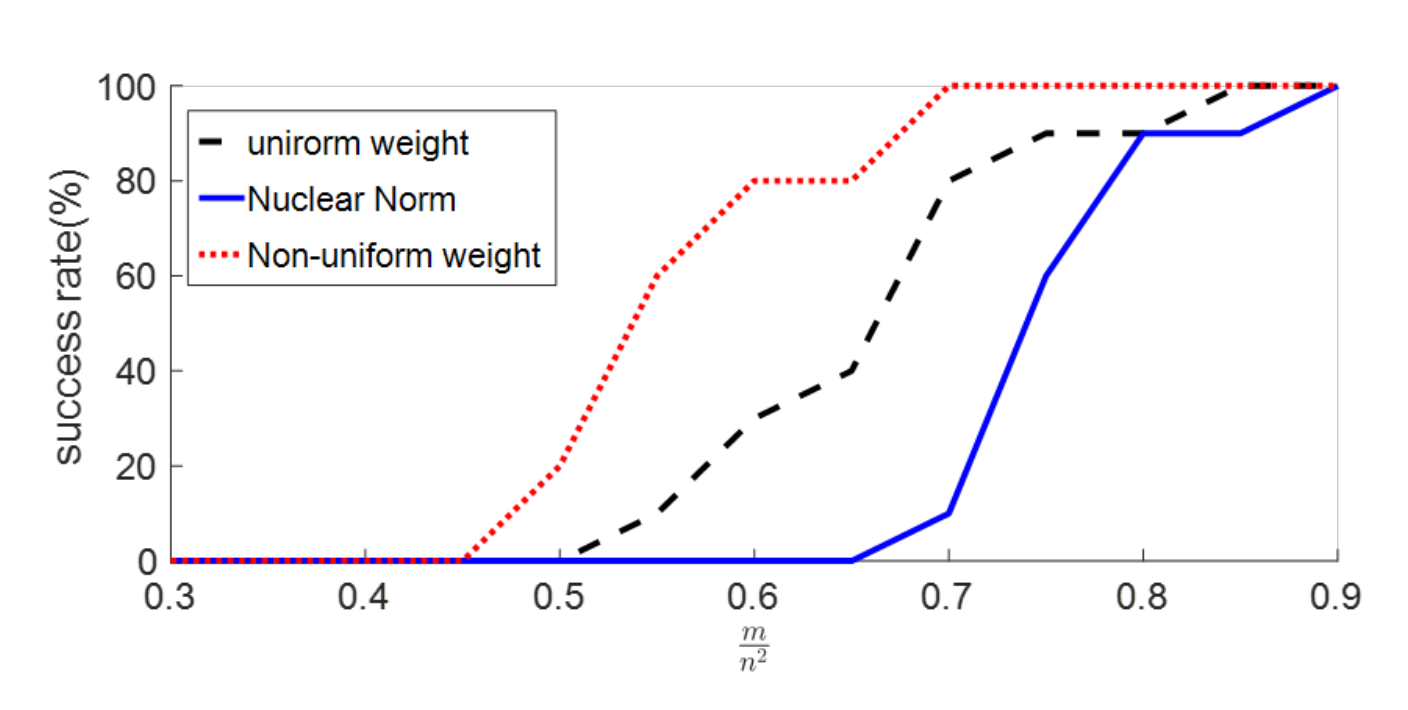}
	\includegraphics[width=3.5in,height=1.5in]{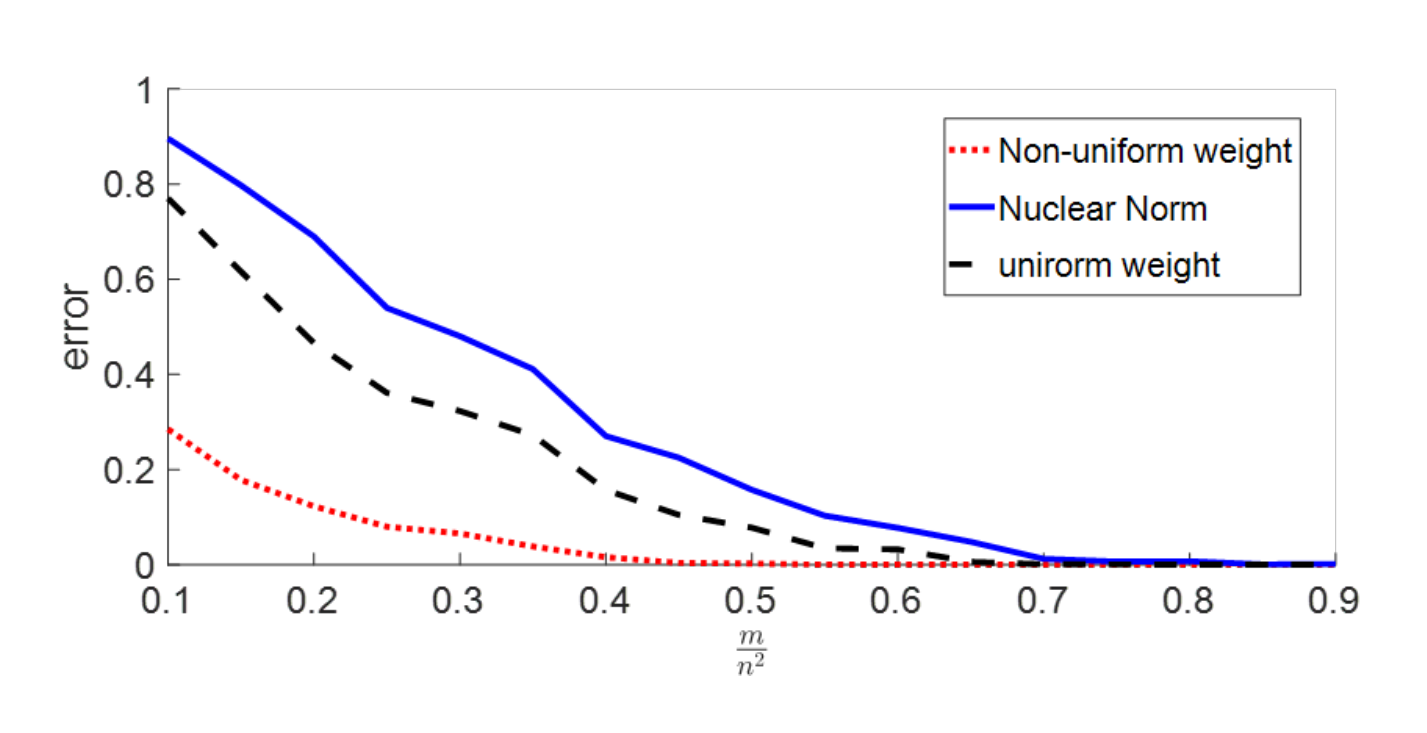}
	\caption{Matrix completion performance using different approaches in the noise-less case. The principal angels are considered as $\bs{\theta}_u = [2.01,8.28,15.55,20.26]$ and $\bs{\theta}_v = [2.09,10.5,19.45,22.00]$.}
 
	\label{Fig1com}
\end{figure*}
\begin{figure*}
	\includegraphics[width=3.5in,height=1.5in]{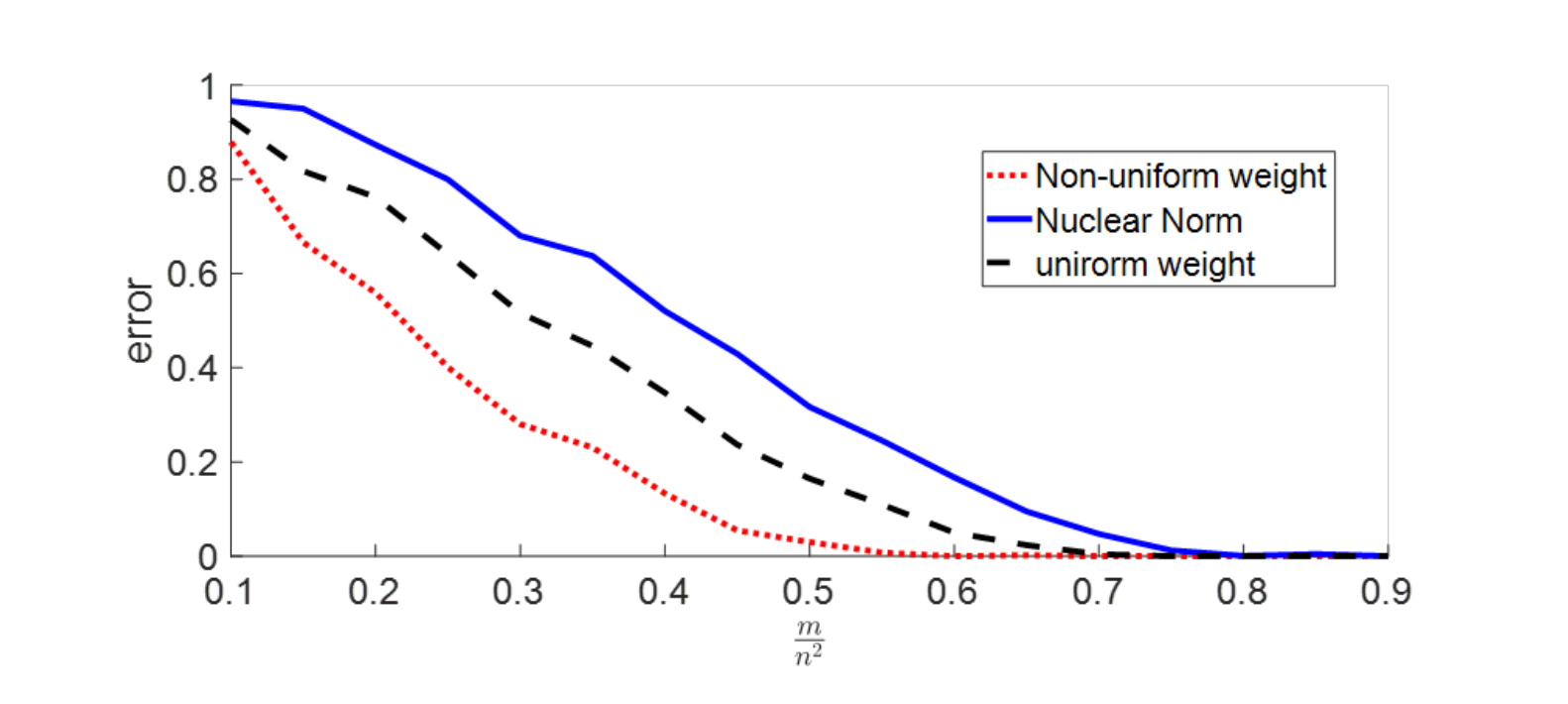}
	\includegraphics[width=3.5in,height=1.5in]{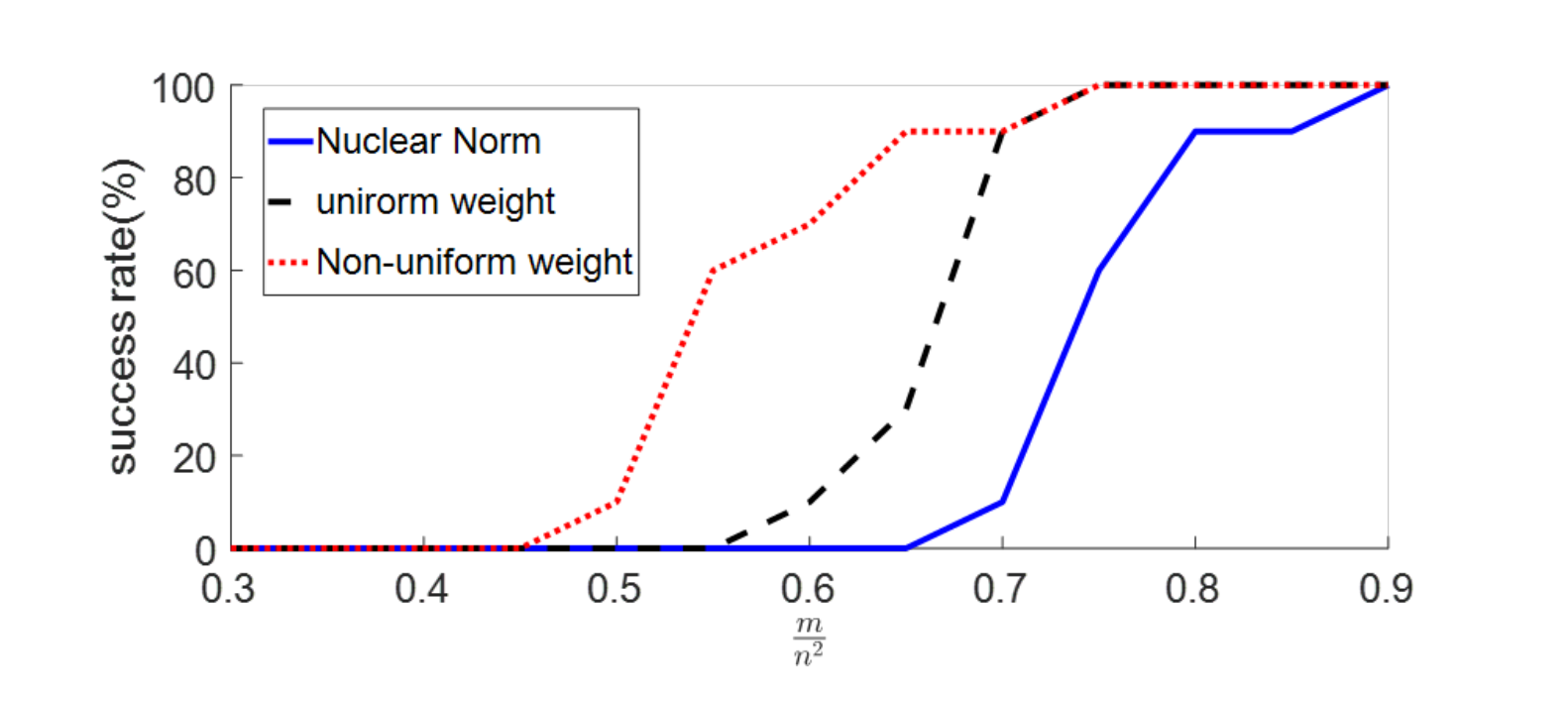}
	\caption{Matrix completion performance using different approaches in the noise-less case. The principal angels are considered as $\bs{\theta}_u = [1.32,1.72,2.11,3.07]$ and $\bs{\theta}_v=[1.08,1.70,2.37,2.73]$.}
	
	\label{Fig2com}
\end{figure*}
\begin{figure*}
	\includegraphics[width=3.5in,height=1.5in]{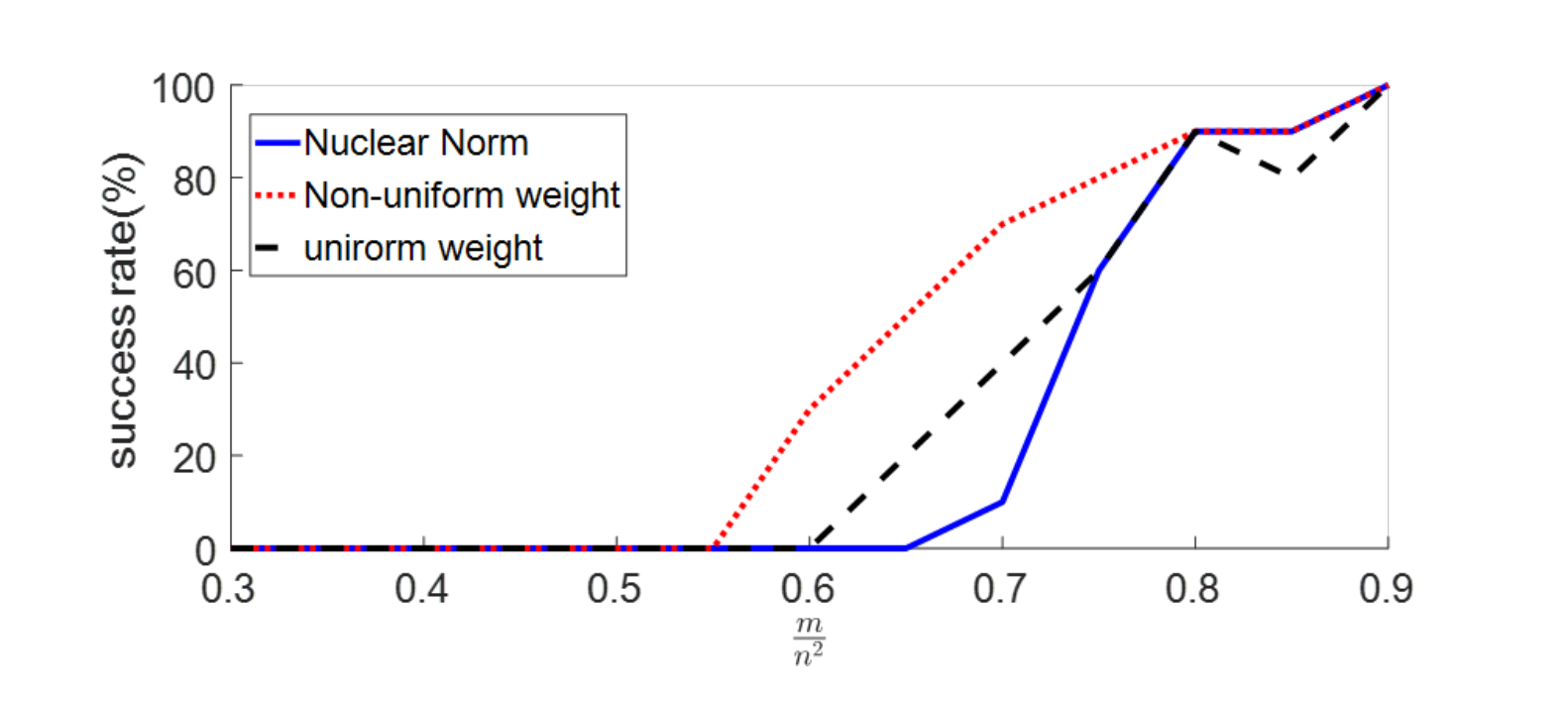}
	\includegraphics[width=3.5in,height=1.5in]{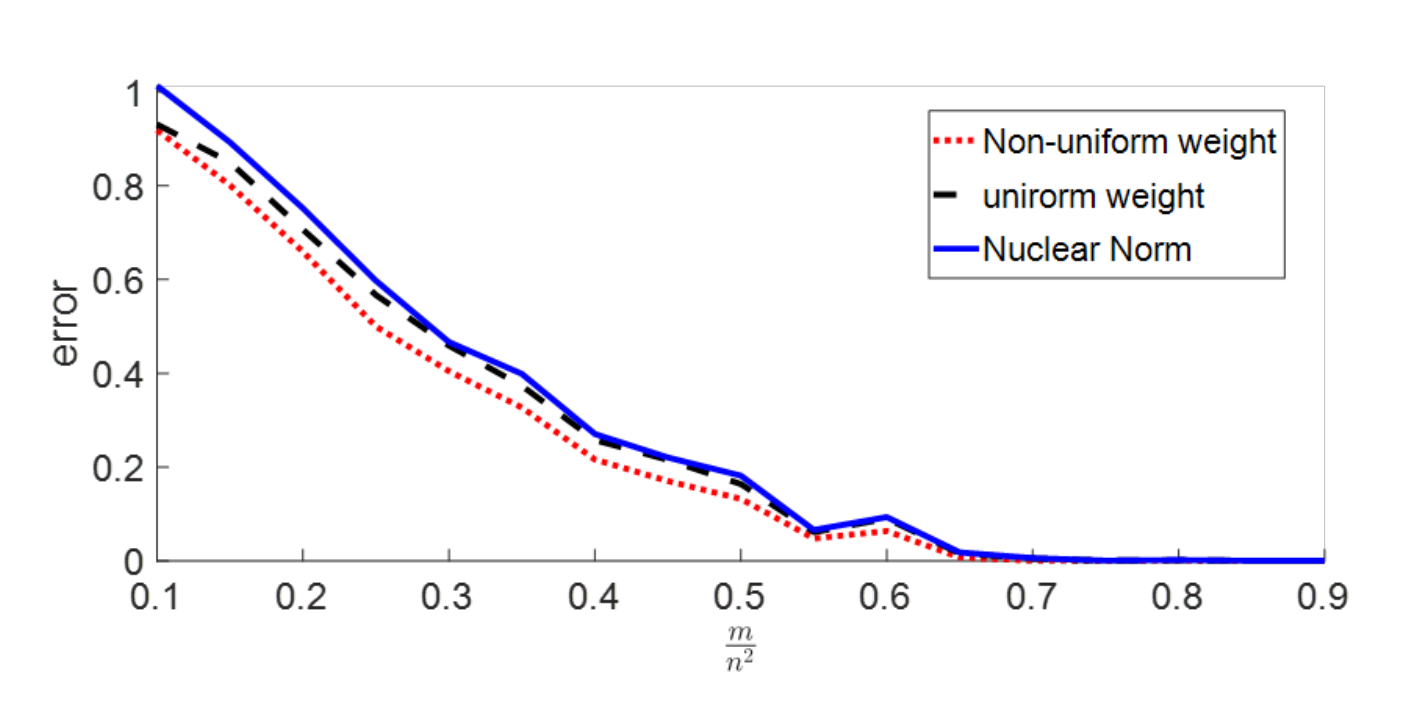}
	\caption{Matrix completion performance using different approaches in the noise-less case. The principal angels are considered as $\bs{\theta}_u = [1.32,1.72,2.11,3.07]$ and $\bs{\theta}_v=[1.08,1.70,2.37,2.73]$.}
	
	\label{Fig3com}
\end{figure*}
In this section, we provide several numerical experiments to verify that the proposed non-uniform weighting strategy performs better than the uniform weighting approach in matrix completion problem with prior subspace information. The optimization problems are solved using CVX package in MATLAB software and the optimal weights are obtained by numerical optimization techniques. Consider a square matrix $\mathbf{ X} \in \mathbb{R}^{ n \times n}$ with $n=20$ and rank $r=4$. We use a perturbation matrix $\mathbf{N} \in \mathbb{R}^{n \times n}$, with independent Gaussian random elements of mean zero and variance $10^{-4}$ to generate $\mathbf{X}'=\mathbf {X} + \mathbf {N}$ with rank $r'=r+4$. Then, we construct the prior subspaces $\widetilde{\bm{\mathcal{U}}}_{r'}$ and $\widetilde{\bm{\mathcal{V}}}_{r'}$ by computing the column and row spaces of $\mathbf {X}'$ and $\mathbf{X}'^H$, respectively. Note that $\bs{\theta}_u \in [0,90]^r$ and $\bs{\theta}_v \in [0,90]^r$ are the known principal angels between subspaces $[\bm{{\mathcal{U}}}_{r} , \widetilde{\bm{\mathcal{U}}}_{r'}]$ and $[\bm{{\mathcal{V}}}_{r} , \widetilde{\bm{\mathcal{V}}}_{r'}]$, respectively. The bases $\mathbf{{U}}_{r}$ and $\mathbf{{U}}_{r'}$ without loss of generality can be defined in such a way that 
\begin{align}
\mathbf{U}_{r}^H\widetilde{\mathbf{U}}_{r'}=[{\rm cos}\bs{\theta}_u~~\mathbf{0}_{r \times r'-r}], \quad 
\mathbf{{V}}_{r}^H\widetilde{\mathbf{V}}_{r'}=[{\rm cos}\bs{\theta}_v~~\mathbf{0}_{r \times r'-r}]
\end{align}
i.e. $\mathbf{{U}}_{r}$ and $\widetilde{\mathbf{U}}_{r'}$ can be considered as the left and right singular matrices of $\mathbf{{U}}_{r}^H\widetilde{\mathbf{U}}_{r'}$. Similar definitions are also done for $\mathbf{{V}}_{r}$ and $\widetilde{\mathbf{V}}_{r'}$.

In this section, we compare the performance of the standard matrix completion \eqref{nuclear_prob} with the strategy of non-uniform weighting \eqref{prior_nuclear_prob}. Also, we examine different angles $\bs{\theta}_u$ and $\bs{\theta}_v$ and compare the proposed strategy with standard matrix completion and single-weight method in \cite{eftekhari2018weighted}. Each experiment is repeated $50$ times with different sampled entries and noise levels (in noisy cases). Considering $\hat{\mathbf {X}}$ as the solution of the problem, the normalized recovery error (NRE) is defined as: ${\rm NRE}\triangleq\tfrac{\| \hat{\mathbf {X}} - \mathbf {X} \|_F}{\| \mathbf {X} \|_F} \nonumber$. NRE values less than $10^{-4}$ is considered to show a successful experiment.

The success rate and NRE (in the noiseless case) are shown in Figure \ref{Fig1com}. In this experiment, prior information is considered to have good accuracy, and the principal angles between subspaces are considered as $\bs{\theta}_u = [1.32,1.72,2.11,3.07]$ and $\bs{\theta}_v=[1.08,1.70,2.37,2.73]$ degrees. As it can be observed, matrix completion with non-uniform weights outperforms the uniform weighting strategy and the unweighted standard matrix completion. 

In Figure \ref{Fig2com}, we investigate a case with similar parameters assuming $\bs{\theta}_u = [2.01,8.28,15.55,20.26]$ and $\bs{\theta}_v = [2.09,10.5,19.45,22.00]$ with the difference that some directions closely align  with the ground-truth ones while others do not. As expected, we see that matrix completion with non-uniform weights performs better than the other methods.

In Figure \ref{Fig3com}, we consider prior subspace information with are far from the ground-truth subspace, i.e. $\bs{\theta}_u = [40.87,49.63,50.55,69.39]$ and $\bs{\theta}_v = [28.76,37.83,40.52,63.65]$. In this experiment, similar to previous ones, it can be again observed that non-uniformly weighting matrix completion has superior performance compared to the other methods.
\section{Conclusion}\label{sec.conclusion}
In this work, we designed a novel framework in order to exploit prior subspace information in matrix completion problem. We first developed a weighted optimization problem to promote both prior subspace knowledge and low-rank feature. Then, we proposed a novel theoretical way to obtain the optimal weights by minimizing the required number of observations. Numerical results were also provided which demonstrate the superior accuracy of our proposed approach compared to the state-of-the-art methods.

\begin{appendices}
\section{Required Lemmas and Backgrounds}\label{sec.required_lemmas}
In this section, some essential lemmas are provided which are required in the proof of Theorem \ref{thm2}.
\subsection{Constructing the Bases}
This section introduces the bases in order to simplify the proofs. 
\begin{lem} \label{Lemma 3-6-1} \cite{daei2018optimal} Consider $\mathbf{X}_r \in \mathbb{R}^{n \times n}$ as a rank $r$ matrix with column and row subspaces $\bm{\mathcal{U}}_r$ and $\bm{\mathcal{V}}_r$, respectively. Also, let $\widetilde{\bm{\mathcal{U}}}_{r'}$ and $\widetilde{\bm{\mathcal{V}}}_{r'}$ of dimension $r'\geq r$ with $r$ known principal angles $\bs{\theta}_u$ and $\bs{\theta}_v$ with subspaces $\bm{\mathcal{U}}_r$ and $\bm{\mathcal{V}}$ as prior information. There exists orthogonal matrices $\mathbf{U}_r , \mathbf{V}_r \in \mathbb{R}^{n \times r}$ and 
$\widetilde{\mathbf{U}}_{r'},\widetilde{\mathbf{V}}_{r'}\in \mathbb{R}^{n \times r'}$
and $\mathbf{B}_L,\mathbf{B}_R \in \mathbb{R}^{n \times n}$ such that:
	\begin{align}
	&\bm{\mathcal{U}}_r = {\rm span}(\mathbf{U}_r), \quad \bm{\mathcal{V}}_r = {\rm span}(\mathbf{V}_r) \nonumber \\
	&\widetilde{\bm{\mathcal{U}}}_{r'} = {\rm span}(\widetilde{\mathbf{U}}_{r'}), \quad \widetilde{\bm{\mathcal{V}}}_{r'}  = {\rm span}(\widetilde{\mathbf{V}}_{r'})\nonumber \\
	&\mathbf{B}_L\triangleq[\mathbf{U}_r~~\mathbf{U}_{1,r}'~~\mathbf{U}_{2,r'-r}' ~~\mathbf{U}_{n-r-r'}'']\in\mathbb{R}^{n\times n} \nonumber \\
	& \mathbf{B}_R\triangleq[\mathbf{V}_r~~ \mathbf{V}_{1,r}' ~~\mathbf{V}_{2,r'-r}'~~ \mathbf{V}_{n-r-r'}'']\in\mathbb{R}^{n\times n}
	\end{align}
	For definitions of the submatrices, see \cite[Section VI.A]{daei2018optimal}.
\end{lem}
The following relation can be concluded from Lemma \ref{Lemma 3-6-1}:
\begin{align}
\widetilde{\mathbf{U}}_{r'}=\mathbf{B}_L
\begin{bmatrix}
{\rm cos}\bs{\theta}_u &  \\ 
-{\rm sin}\bs{\theta}_u &  \\
& -\mathbf{I}_{r'-r} \\
&  \\
\end{bmatrix}.
\end{align}
Therefore, orthogonal projections onto the subspaces $\widetilde{\bm{\mathcal U}}_{r'}$ and $\widetilde{\bm{\mathcal U}}_{r'}^{\perp}$ are:
\begin{align}
&\mathbf{P}_{\widetilde{\bm{\mathcal U}}_{r'}}=\widetilde{\mathbf{U}}_{r'}\widetilde{\mathbf{U}}_{r'}^{H}=\nonumber \\
&\mathbf{B}_L 
\begin{bmatrix}
{\rm cos}^2\bs{\theta}_u & -{\rm sin}\bs{\theta}_u {\rm cos}\bs{\theta}_u &  &\\
-{\rm sin}\bs{\theta}_u {\rm cos}\bs{\theta}_u & {\rm sin}^2\bs{\theta}_u &  &\\
&  & \mathbf{I}_{r'-r} &\\
&&&
\end{bmatrix}
\mathbf{B}_L ^H \nonumber\\
&\mathbf{P}_{\widetilde{\bm{\mathcal {U}}}_{r'}^{\perp}}=\mathbf{I} - \mathbf{P}_{\widetilde{\bm{\mathcal U}}_{r'}}= \nonumber \\
&\mathbf{B}_L 
\begin{bmatrix}
{\rm sin}^2\bs{\theta}_u & {\rm sin}\bs{\theta}_u {\rm cos}\bs{\theta}_u &  &\\
{\rm sin}\bs{\theta}_u {\rm cos}\bs{\theta}_u & {\rm cos}^2\bs{\theta}_u &  &\\
&  & \mathbf{0}_{r'-r} &\\
&&&\mathbf{I}_{n-r'-r}
\end{bmatrix}\mathbf{B}_L ^H
\end{align}
We also have:
\begin{align}\label{34}
&\mathbf{Q}_{\widetilde{\bm{\mathcal{U}}}_{r^{\prime}}} \triangleq \widetilde{\mathbf{U}}_{r^{\prime}}\bs{\Lambda}\widetilde{\mathbf{U}}_{r^{\prime}}^{\rm{H}} + \mathbf{P}_{\widetilde{\bm{\mathcal{U}}}^{\perp}}= \nonumber \\
& \mathbf{B}_L\scalebox{.8}{$\begin{bmatrix}
  \bs{\Lambda}_1 {\rm cos}^2\bs{\theta}_u + {\rm sin}^2\bs{\theta}_u&(\mathbf{I}-\bs{\Lambda}_1) {\rm sin}\bs{\theta}_u {\rm cos}\bs{\theta}_u&&\\
  (\mathbf{I}-\bs{\Lambda}_1) {\rm sin}\bs{\theta}_u {\rm cos}\bs{\theta}_u&\bs{\Lambda}_1 {\rm sin}^2\bs{\theta}_u+ {\rm cos}^2\bs{\theta}_u&&\\
  &\bs{\Lambda}_2&  \\
&  &\mathbf{I}_{n-r^{\prime}-r}
\end{bmatrix}$}\mathbf{B}_L^{\rm H}
\end{align}	
in which $ \bs{\Lambda} \triangleq \begin{bmatrix}
\bs{\Lambda}_1 \in \mathbb{R}^{r \times r}& \\
& \bs{\Lambda}_2 \in \mathbb{R}^{r^{\prime}-r \times r^{\prime}-r}
\end{bmatrix} $.

Define
\begin{align}
   \mathbf{O}_L \triangleq \scalebox{.7}{$\begin{bmatrix}
        (\bs{\Lambda}_1 {\rm cos}^2\bs{\theta}_u + {\rm sin}^2\bs{\theta}_u) \cdot \bs{\Delta}_L^{-1}&-(\mathbf{I}-\bs{\Lambda}_1) {\rm sin}\bs{\theta}_u {\rm cos}\bs{\theta}_u \cdot \bs{\Delta}_L^{-1}&&\\
        -(\mathbf{I}-\bs{\Lambda}_1) {\rm sin}\bs{\theta}_u {\rm cos}\bs{\theta}_u \cdot \bs{\Delta}_L^{-1}&(\bs{\Lambda}_1 {\rm cos}^2\bs{\theta}_u + {\rm sin}^2\bs{\theta}_u \cdot \bs{\Delta}_L^{-1}&\\
        & \mathbf{I}_{r^{\prime}-r} &   \\
&  & \mathbf{I}_{n-r^{\prime}-r}
    \end{bmatrix},$}
\end{align}

in which $ \bs{\Lambda}_1 \succeq \mathbf{0} $, and $ \bs{\Delta}_L \triangleq \sqrt{\bs{\Lambda}_1^2 {\rm cos}^2\bs{\theta}_u + {\rm sin}^2\bs{\theta}_u}\in \mathbb{R}^{n \times n }$  is an invertible matrix. Now \eqref{34} can be rewritten as:
\begin{align}
&\mathbf{Q}_{\widetilde{\bm{\mathcal{U}}}_{r^{\prime}}} =\mathbf{B}_L (\mathbf{O}_L\mathbf{O}_L^{\rm H})
\scalebox{.7}{$\begin{bmatrix}
  \bs{\Lambda}_1 {\rm cos}^2\bs{\theta}_u + {\rm sin}^2\bs{\theta}_u&(\mathbf{I}-\bs{\Lambda}_1) {\rm sin}\bs{\theta}_u {\rm cos}\bs{\theta}_u&&\\
  (\mathbf{I}-\bs{\Lambda}_1) {\rm sin}\bs{\theta}_u {\rm cos}\bs{\theta}_u&\bs{\Lambda}_1 {\rm sin}^2\bs{\theta}_u+ {\rm cos}^2\bs{\theta}_u&&\\
  &\bs{\Lambda}_2&  \\
&  &\mathbf{I}_{n-r^{\prime}-r}
\end{bmatrix}$}
\mathbf{B}_L^{\rm H} \nonumber \\
&=\mathbf{B}_L \mathbf{O}_L 
\scalebox{.9}{$
\begin{bmatrix}
    \bs{\Delta}_L&(\mathbf{I}-\bs{\Lambda}_{1}^{2}) {\rm sin}\bs{\theta}_u {\rm cos}\bs{\theta}_u \cdot\bs{\Delta}_L^{-1}&&\\
   & \bs{\Lambda}_1\bs{\Delta}_L^{-1} & &\\
   &&\bs{\Lambda}_2&  \\
&  &&\mathbf I_{n-r'-r}
\end{bmatrix}$}
\mathbf{B}_L^{\rm H} \nonumber \\
& =: \mathbf{B}_L \mathbf{O}_L \begin{bmatrix}
\mathbf{L}_{11} & \mathbf{L}_{12} & & \\
&\mathbf{L}_{22}& & \\ 
&&\bs{\Lambda}_2&\\
& & &\mathbf{I}_{n-r^{\prime}-r}
\end{bmatrix} \mathbf{B}_L^{\rm H}  \nonumber \\
& = \mathbf{B}_L \mathbf{O}_L \mathbf{L} \mathbf{B}_L^{\rm H}, 
\end{align}
where $\mathbf{L} \in \mathbb{R}^{n \times n }$ is  an upper-triangular block matrix:
\begin{align}\label{37}
&\mathbf{L} \triangleq\begin{bmatrix}
\mathbf{L}_{11} & \mathbf{L}_{12} & & \\
& \mathbf{L}_{22} & & \\
&&\bs{\Lambda}_2&\\
& & &\mathbf{I}_{n-r^{\prime}-r}
\end{bmatrix}  \nonumber \\
&=\begin{bmatrix}
\bs{\Delta}_L  &  (\mathbf{I}-\bs{\Lambda}_{1}^{2}) {\rm sin}\bs{\theta}_u {\rm cos}\bs{\theta}_u\cdot \bs{\Delta}_L^{-1}  & & \\
& \bs{\Lambda}_1\bs{\Delta}_L^{-1} &  & \\
& & \bs{\Lambda}_2& \\
& & &\mathbf{I}_{n-r^{\prime}-r} \\
\end{bmatrix}.
\end{align}
Since matrices $\mathbf{B}_L$ and $\mathbf{O}_L$ are orthonormal bases, it follows that $\| \mathbf{Q}_{\widetilde{\bm{\mathcal{U}}}_{r^{\prime}}} \| = \| \mathbf{L} \| = 1$.
Similar results can  also be deduced for the row subspace:
\begin{align}
&\mathbf{R} \triangleq\begin{bmatrix}
\mathbf{R}_{11} & \mathbf{R}_{12} & & \\
& \mathbf{R}_{22} & & \\
& &\bs{\Gamma}_2 & \\
& & & \mathbf{I}_{n-r^{\prime}-r}
\end{bmatrix}  \nonumber \\
&= \begin{bmatrix}
\bs{\Delta}_R  &  (\mathbf{I}-\bs{\Gamma}_{1}^{2}) {\rm sin}\bs{\theta}_v {\rm cos}\bs{\theta}_v\cdot\bs{\Delta}_R^{-1}  & &  \\
& \bs{\Gamma}_{1}\bs{\Delta}_R^{-1} & & \\
& & \bs{\Gamma}_2 & \\
& & & \mathbf{I}_{n-r^{\prime}-r}  
\end{bmatrix},
\end{align}
where $ \mathbf{\Delta}_R \triangleq \sqrt{\bs{\Gamma}_{1}^{2} {\rm cos}^2\bs{\theta}_v + {\rm sin}^2\bs{\theta}_v} $ and  $ \bs{\Delta}_L $ is similarly defined. Considering $ \mathbf{Z} \in \mathbb{R}^{n \times n } $  as an arbitrary matrix, one may write:
\begin{align}\label{40}
&\mathbf{Q}_{\widetilde{\bm{\mathcal{U}}}_{r^{\prime}}}\mx{Z}\mathbf{Q}_{\widetilde{\bm{\mathcal{V}}}_{r^{\prime}}} = \mathbf{B}_L \mathbf{O}_L \mathbf{L} (\mathbf{B}_L^{\rm H} \mx{Z} \mathbf{B}_R) \mathbf{R}^{\rm H} \mathbf{O}_R^{\rm H}  \mathbf{B}_R^{\rm H}   \nonumber \\
&= \mathbf{B}_L \mathbf{O}_L \mathbf{L} \overline{\mx{Z}} \mathbf{R}^{\rm H} \mathbf{O}_R^{\rm H} \mathbf{B}_R^{\rm H} \quad (\overline{\mx{Z}}\triangleq\mathbf{B}_L^{\rm H} \mx{Z} \mathbf{B}_R )  \nonumber \\
&\triangleq\mathbf{B}_L \mathbf{O}_L \mathbf{L}
\begin{bmatrix}
\overline{\mx{Z}}_{11} & \overline{\mx{Z}}_{12} &\overline{\mx{Z}}_{13}& \overline{\mx{Z}}_{14}\\
\overline{\mx{Z}}_{21} & \overline{\mx{Z}}_{22} & \overline{\mx{Z}}_{23} & \overline{\mx{Z}}_{24}\\
\overline{\mx{Z}}_{31} & \overline{\mx{Z}}_{32}&\overline{\mx{Z}}_{33}&\overline{\mx{Z}}_{34}\\
\overline{\mx{Z}}_{41} &  \overline{\mx{Z}}_{42} & \overline{\mx{Z}}_{43}& \overline{\mx{Z}}_{44}
\end{bmatrix}\mathbf{R}^{\rm H} \mathbf{O}_R^{\rm H} \mathbf{B}_R^{\rm H}.
\end{align}
Since ${\rm span}(\mathbf{X}_r) = {\rm span}(\mathbf{U}_r)$ and ${\rm span}(\mathbf{X}_r^{\rm H}) = {\rm span}(\mathbf{V}_r)$ and with upper triangular matrices $\mathbf{L}$ and $\mathbf{R}$, we can rewrite $ \mathbf{Q}_{\widetilde{\bm{\mathcal{U}}}_{r^{\prime}}}  \mathbf{X}_r \mathbf{Q}_{\widetilde{\bm{\mathcal{V}}}_{r^{\prime}}} $ in terms of new bases as follows:
\begin{align}\label{42}
&\mathbf{Q}_{\widetilde{\bm{\mathcal{U}}}_{r^{\prime}}}  \mathbf{X}_r \mathbf{Q}_{\widetilde{\bm{\mathcal{V}}}_{r^{\prime}}} = \mathbf{B}_L \mathbf{O}_L \mathbf{L} (\mathbf{B}_L^{\rm H} \mathbf{X}_r \mathbf{B}_R) \mathbf{R}^{\rm H} \mathbf{O}_R^{\rm H}  \mathbf{B}_R^{\rm H}   \nonumber \\
&= \mathbf{B}_L \mathbf{O}_L \mathbf{L} \overline{\mathbf{X}}_r \mathbf{R}^{\rm H} \mathbf{O}_R^{\rm H} \mathbf{B}_R^{\rm H} \quad (\overline{\mathbf{X}}_r\triangleq\mathbf{B}_L^{\rm H} \mathbf{X}_r \mathbf{B}_R )  \nonumber \\
&\triangleq \mathbf{B}_L \mathbf{O}_L \mathbf{L}
\begin{bmatrix} 
\overline{\mathbf{X}}_{r,11} &  \\
&  \mathbf{0}_{n-r} 
\end{bmatrix}\mathbf{R}^{\rm H} \mathbf{O}_R^{\rm H} \mathbf{B}_R^{\rm H} \nonumber \\ 
&=\mathbf{B}_L \mathbf{O}_L 
\begin{bmatrix}
\mathbf{L}_{11}\overline{\mathbf{X}}_{r,11}\mathbf{R}_{11} &  \\
&  \mathbf{0}_{n-r} 
\end{bmatrix}\mathbf{O}_R^{\rm H} \mathbf{B}_R^{\rm H}.
\end{align}
\begin{lem}\label{lem 4}
	The operator norms regarding the sub-blocks of $\mx{L}$ in \eqref{37} and \eqref{def_fun} are as follows:
	\begin{align}
&\|\mathbf{L}_{11}\| = \|\bs{\Delta}_L\| 
=\sqrt{\max_{i} \{f^2_2(\lambda_1(i) , \theta_u(i))}\}, \nonumber \\ 
& \|\mathbf{L}_{12}\| 
=\sqrt{\max_i \Big\{\tfrac{f^2_2(1-\lambda^2_{1}(i) , \theta_u(i))}{f^2_2(\lambda_1(i) , \theta_u(i))}}\Big\},  \nonumber \\
& \|\mathbf{I}_r - \mathbf{L}_{22} \| 
=\sqrt{\max_i \Big\{\Big(\tfrac{\lambda_{1}(i)-f_2(\lambda_1(i) , \theta_u(i))}{f_2(\lambda_1(i) , \theta_u(i))}\Big)^2\Big\}},  \nonumber \\
& \label{eq:L11L12}\|[\mathbf{L}_{11} \quad  \mathbf{L}_{12}]\| 
=\max_i \Big\{ \tfrac{f_1(\lambda_1(i) , \theta_u(i))}{f_2(\lambda_1(i) , \theta_u(i))}\Big\}  \\ 
& \label{eq:L'} \|\mathbf L ^ \prime \|^{2} = \max_i \{{d}_i(\bs{\theta}_{u},\mathbf \lambda_{1},\mathbf \lambda_{2})\ \\
& \Big\|\begin{bmatrix}
\mathbf{I}_{r} - \mathbf{L}_{22} & \\
& \mathbf{I}_{r} - \bs{\Lambda}_{2}
\end{bmatrix} \Big\| 
=\nonumber\\
&\max \Big \{  \max_i \{1-\lambda_{2}(i)\}, \max_i \{1 - \tfrac{\lambda_{1}(i)}{f_2(\lambda_1(i) , \theta_u(i))}\} \Big \}, 
	\end{align} 
	where $d_1, d_2$ are defined as:
	\begin{align}
&d_1(\bs{\theta},\mathbf a,\mathbf b)\triangleq \nonumber \\
&\underset{i}{\rm max} \bigg\{ \big( \tfrac{a(i)}{f_2(a(i),\theta(i))} -1 \big)^2 + \tfrac{(1-a(i))^2\cos^2\theta(i)\sin^2\theta(i)}{a(i)^2\cos^2\theta(i)\sin^2\theta(i)} \bigg\} \nonumber \\
& d_2(\bs{\theta},\mathbf a,\mathbf b)\triangleq\underset{i}{\rm max}(b(i)-1)^2
	\end{align}
	The same relations hold for the sub-blocks of $ \mathbf{R} $. 
	
	Proof.  See Appendix \ref{proof lemma 4}.
\end{lem}
\subsection{Support Definitions}
Let $\mathbf{X}_r \in \mathbb{R}^{n \times n}$ be a rank-$r$ matrix which is obtained via the truncated SVD of $ \mathbf{X} $: 
$$ \mathbf{X} = \mathbf{X}_r + \mathbf{X}_{r^+} = \mathbf{U}_r\overline{\mathbf{X}}_{r,11}\mathbf{V}_r^{\rm H}+\mathbf{X}_{r^+},$$
where $ \mathbf{U}_r $ and $ \mathbf{V}_r $ are some orthogonal bases of column and row spaces of $\mathbf{X}_r$, and thus $ \overline{\mathbf{X}}_{r,11} $ is not necessarily diagonal. Also consider that  $ \bm{\mathcal{U}}_{r} = {\rm span}(\mathbf{U}_r)= {\rm span}(\mathbf{X}_r) $ and $ \bm{\mathcal{V}}_{r} = {\rm span}(\mathbf{V}_r)= {\rm span}(\mathbf{X}_r^{\rm H}) $ are column and row subspaces of $  \mathbf{X}_r$, respectively. Then the support of $ \mathbf{X}_r $ can be defined as:  
\begin{align}\label{44}
&\mathbf{T} \triangleq \{\mathbf{Z}\in\mathbb{R}^{n\times n} : \mathbf{Z}=\mathbf{P}_{\bm{\mathcal{U}}_r}\mathbf{Z}\mathbf{P}_{\bm{\mathcal{V}}_r}+\mathbf{P}_{\bm{\mathcal{U}}_r}\mathbf{Z}\mathbf{P}_{\bm{\mathcal{V}}_r^\perp} \nonumber \\
&\quad \quad \quad \quad \quad \quad 	+\mathbf{P}_{\bm{\mathcal{U}}_r^\perp}\mathbf{Z}\mathbf{P}_{\bm{\mathcal{V}}_r^\perp}\}   = {\rm supp}(\mathbf{X}_r),
\end{align}
and the orthogonal projection onto $\mathbf{T}$ and $\mathbf{T}^{\perp}$ can be defined as
\begin{align}
&\mathcal{P}_{\mathbf{T}}(\mathbf{Z}) = \mathbf{P}_{\bm{\mathcal{U}}}\mathbf{Z} +\mathbf{Z}\mathbf{P}_{\bm{\mathcal{V}}} - \mathbf{P}_{\bm{\mathcal{U}}}\mathbf{Z}\mathbf{P}_{\bm{\mathcal{V}}},\nonumber \\
&  \mathcal{P}_{\mathbf{T}^{\perp}}(\mathbf{Z}) =  \mathbf{P}_{\bm{\mathcal{U}}^{\perp}}\mathbf{Z}\mathbf{P}_{\bm{\mathcal{V}}^{\perp }}.
\end{align} 
We can rewrite $ \mathbf{T} $  using Lemma \ref{Lemma 3-6-1} as 
\begin{align} \label{3-42}
&\mathbf{T} = \Big\{\mathbf{Z}\in\mathbb{R}^{ n\times n} : \mathbf{Z}=\mathbf{B}_L \overline{\mathbf{Z}} \mathbf{B}_R^{\rm H}, \quad \overline{\mathbf{Z}}\triangleq
\begin{bmatrix}
\overline{\mathbf{Z}}_{11} & \overline{\mathbf{Z}}_{12} \\
\overline{\mathbf{Z}}_{21} & \mathbf{0}_{n-r}
\end{bmatrix} \Big\}  \nonumber \\
&=\mathbf{B}_L \overline{\mathbf{T}} \mathbf{B}_R^{\rm H},
\end{align}
where $ \overline{\mathbf{T}} \subset \mathbb{R}^{n \times n} $  is the support of $ \overline{\mathbf{X}}_r = \mathbf{B}_L^{\rm H} \mathbf{X}_r \mathbf{B}_R $:
\begin{align}
\overline{\mathbf{T}} = \{\overline{\mathbf{Z}}\in\mathbb{R}^{ m\times n} : \overline{\mathbf{Z}}\triangleq	\begin{bmatrix}
\overline{\mathbf{Z}}_{11} & \overline{\mathbf{Z}}_{12} \\
\overline{\mathbf{Z}}_{21} & \mathbf{0}_{n-r}
\end{bmatrix}\}.
\end{align}
For arbitrary 
\begin{align}
\overline{\mathbf{Z}}\triangleq	\begin{bmatrix}
\overline{\mathbf{Z}}_{11} & \overline{\mathbf{Z}}_{12} \\
\overline{\mathbf{Z}}_{21} & \overline{\mathbf{Z}}_{22}
\end{bmatrix} \in \mathbb{R}^{n\times n},
\end{align}
the orthogonal projection onto $ \overline{\mathbf{T}} $ and its complement $ \overline{\mathbf{T}}^{\perp} $ are  
\begin{align}\label{47}
\mathcal{P}_{\overline{\mathbf{T}}}(\overline{\mathbf{Z}}) = \begin{bmatrix}
\overline{\mathbf{Z}}_{11} & \overline{\mathbf{Z}}_{12} \\
\overline{\mathbf{Z}}_{21} & \mathbf{0}_{n-r}
\end{bmatrix}, \quad \mathcal{P}_{\overline{\mathbf{T}}^{\perp}} (\overline{\mathbf{Z}})= \begin{bmatrix}
\mathbf{0}_{r} & \\
& \overline{\mathbf{Z}}_{22}
\end{bmatrix}, 
\end{align}
respectively. Since $ \mathbf{Z} = \mathbf{B}_L \overline{\mathbf{Z}} \mathbf{B}_R^{\rm H}  $, one can say:   
\begin{align}\label{48}
& \mathcal{P}_{\mathbf{T}}(\mathbf{Z}) = \mathbf{B}_L \mathcal{P}_{\overline{\mathbf{T}}}(\overline{\mathbf{Z}})\mathbf{B}_R^{H}, \quad
\mathcal{P}_{\mathbf{T}^{\perp}}(\mathbf{Z}) = \mathbf{B}_L \mathcal{P}_{\overline{\mathbf{T}}^{\perp}}(\overline{\mathbf{Z}}) \mathbf{B}_R^{H}.
\end{align}

\section{Proof of Theorem \ref{thm2}}\label{proof_th}
\begin{proof} 
For matrix $\mathbf{X}_r$ with rank $r$, if $\mathbf{U}_r$ and $\mathbf{V}_r$ indicate the orthogonal bases, then the column and row subspaces are:
$$\bm{\mathcal{U}}_r = {\rm span}(\mathbf{X}_r) = {\rm span}(\mathbf{U}_r) , \bm{\mathcal{V}}_r = {\rm span}(\mathbf{X}_r^H) = {\rm span}(\mathbf{V}_r).$$
The $i$th and $l$-th coherence parameters of the column and row spaces of $\mathbf{X}_r$ can be expressed as 
\begin{align}
&\mu_i = \mu_i(\bm{\mathcal{U}}_r) \triangleq \tfrac{n}{r} \| \mathbf{U}_r[i,:] \|_{2}^2  \quad i \in [1:n],  \nonumber \\
&\nu_l = \nu_l(\bm{\mathcal{V}}_r) \triangleq \tfrac{n}{r} \| \mathbf{V}_r[l,:] \|_{2}^2  \quad l \in [1:n].
\end{align}	
As we can see in these definitions, the coherence parameter of subspaces is independent of the selection of orthogonal bases. According to the definition \ref{coherence} and \eqref{eq:coherence_parameter_of_matrix}, the coherence parameter of a matrix is equal to the maximum coherence of its column and row subspaces.
Similarly, for subspaces $\breve{\bm{\mathcal{U}}}_r={\rm span}([\mathbf{U}_r , \widetilde{\mathbf{U}}_{r'}])$ and $\breve{\bm{\mathcal{V}}}_r={\rm span}([\mathbf{V}_r , \widetilde{\mathbf V}_{r'}]) $ we have:
\begin{align}
\breve{\mu}_i = \breve{\mu}_i(\breve{\bm{\mathcal{U}}}_r)  \quad i \in [1:n] \quad
\breve{\nu}_l = \nu_l(\breve{\bm{\mathcal{V}}}_r)  \quad l \in [1:n]
\end{align}
For simplicity and in order to use coherence between subspaces, we define the following diagonal matrix:
\begin{align}
\bs{\mu}\triangleq  \begin{bmatrix}
\mu_{1} & & \\
& \ddots & \\
& & \mu_{n}
\end{bmatrix}.
\end{align}
Let $\| \mathbf A \|_{2 \rightarrow \infty }$ be the maximum $\ell_2$-norm of the rows in $\mathbf A$, then, the following relations hold:
\begin{align}
&\|(\tfrac{\mu r}{n})^{\tfrac{-1}{2}} \mathbf{U}_r \|_{2 \rightarrow \infty} =1, \|(\tfrac{\nu r}{n})^{\tfrac{-1}{2}} \mathbf{V}_r \|_{2 \rightarrow \infty} =1  \nonumber \\
&\|(\tfrac{\mu r}{n})^{\tfrac{-1}{2}} \mathbf{U}^{\prime}_{r^{\prime}} \|_{2 \rightarrow \infty} \le \sqrt{\tfrac{r^{\prime}+r}{r}\max_{i}\tfrac{\breve{\mu}_i}{\mu_i}}  \nonumber \\
&\|(\tfrac{\nu r}{n})^{\tfrac{-1}{2}} \mathbf{V}^{\prime}_{r^{\prime}} \|_{2 \rightarrow \infty} \le \sqrt{\tfrac{r^{\prime}+r}{r}\max_{l}\tfrac{\breve{\nu}_l}{\nu_l}}
\end{align}
By considering $\breve{\mathbf {U}}$ as the orthogonal basis of $\breve{\bm{\mathcal{U}}}$, it holds that:
\begin{align}
&\|(\tfrac{\mu r}{n})^{\tfrac{-1}{2}} \mathbf{U}^{\prime}_r \|_{2 \rightarrow \infty} = \max_{i} \tfrac{\| \mathbf{U}^{\prime}_r[i,:]\|_{2}}{\| \mathbf{U}_r[i,:]\|_{2}}  \nonumber \\
&\le \max_{i}\tfrac{\| \breve{\mathbf{U}}_r[i,:]\|_{2}}{\| \mathbf{U}_r[i,:]\|_{2}} = \sqrt{\max_{i}\tfrac{\breve{\mu}_i \cdot {\rm dim}(\breve{\mathbf{U}}_r)}{\mu_i r}}  \nonumber \\
& \le \sqrt{\tfrac{r^{\prime}+r}{r} \max_{i}\tfrac{\breve{\mu}_i}{\mu_i}} \quad ({\rm dim}(\breve{\mathbf{U}}_r) \le r^{\prime}+r)  \label{proof5}
\end{align} 
where the third inequality of \eqref{proof5} comes from the fact that $\bm{\mathcal{U}_{r'}'}\subset \breve{\bm{\mathcal{U}}}$.

Now, we provide some properties of sampling operator \eqref{sampling_opr}, which is necessary for developing our main theory. In order to list these properties, we initially define the following norm functions. Assume $\mu(\infty)$ and $\mu(\infty,2)$ describe the weighted $\ell_{\infty}$-norm and the maximum weighted $\ell_2$-norm of rows of a matrix, respectively. For a square matrix $\mathbf Z \in \mathbb{R}^{n \times n}$ we set:
\begin{align}
&\| \mathbf{Z} \|_{\mu(\infty)} = \| (\tfrac{\mu r}{n})^{\tfrac{-1}{2}} \mathbf{Z} (\tfrac{\nu r}{n})^{\tfrac{-1}{2}} \| _{\infty}  =\max_{i,j}\sqrt{\tfrac{n}{\mu_i r}}|\mathbf{Z}[i,l]|\sqrt{\tfrac{n}{\nu_l r}}
\end{align}
\begin{align}
&\| \mathbf{Z} \|_{\mu(\infty,2)} = \| (\tfrac{\mu r}{n})^{\tfrac{-1}{2}} \mathbf{Z}\|_{2 \rightarrow \infty} \vee \|(\tfrac{\nu r}{n})^{\tfrac{-1}{2}} \mathbf{Z}^{\rm H}\| _{2 \rightarrow \infty}  \nonumber \\
&=\max_{i} \{\sqrt{\tfrac{n}{\mu_i r}}\|\mathbf{Z}[i,:]\|_2\}\vee \max_{l}\{\sqrt{\tfrac{n}{\nu_l r}} \|\mathbf{Z}[:,l]\|_2\}
\end{align} 
\begin{lem}{\cite{chen2015completing}} \label{Lemma1}
	Given the sampling operator as defined in \eqref{sampling_opr} and assuming $\mathbf{T}$ as the support of $\mx{X}_r$ and $\mathcal{P}_\mathbf{T}$ as the orthogonal projection onto the support, the following inequality holds with high probability: 
$$ \|(\mathcal{P}_{\mathbf{T}} - \mathcal{P}_{\mathbf{T}}\circ \mathcal{R} _\Omega\circ\mathcal{P}_{\mathbf{T}})(\mathbf{Z}) \|_{F \rightarrow F} \le \tfrac{1}{2}$$
under the condition that
$$ \tfrac{(\mu_i +\nu_l)r \log{n}}{n} \le p_{il} \le 1 \quad \forall i,l \in [1:n].$$
\end{lem}
Above, $ \|\mathcal{A}(\cdot)\|_{F \rightarrow F} \triangleq  {\rm sup}_{\|\bm{X}\|_F \le 1 } \|\mathcal{A}(\bm{X})\|_F$ is the operator norm of the linear map $\mathcal{A}(\cdot)$, and $ (\mathcal{A} \circ \mathcal{B})(\cdot) = \mathcal{A}(\mathcal{B}(\cdot))$ stands for the composition of operators $ \mathcal{A}$ and  $ \mathcal{B}$.

\begin{lem}{\cite{chen2015completing}} Given the assumptions stated in Lemma \ref{Lemma1}, for any matrix $\mathbf{Z}$,  with high probability, we expect 	
	$$ \|(\mathcal{I} - \mathcal{R} _\Omega)(\mathbf{Z}) \| \le \| \mathbf{Z} \|_{\mu(\infty)} + \| \mathbf{Z} \|_{\mu(\infty,2)} $$
	subject to the conditions stated in Lemma \ref{Lemma1}.
\end{lem}

\begin{lem}
	\cite{chen2015completing} 
Given the assumptions in Lemma \ref{Lemma1}, for any matrix $\mathbf {Z}$ such that $\mathcal{P} (\mathbf {Z}) = \mathbf {Z}$, we expect
	\begin{align*}
	\|(\mathcal{P}_{\mathbf{T}} - \mathcal{P}_{\mathbf{T}}\circ \mathcal{R} _\Omega\circ\mathcal{P}_{\mathbf{T}})(\mathbf{Z}) \|_{\mu(\infty,2)}\le \tfrac{1}{2}\| \mathbf{Z} \|_{\mu(\infty,2)} + \tfrac{1}{2}\| \mathbf{Z} \|_{\mu(\infty)}
	\end{align*}
	with high probability under the same condition as Lemma \ref{Lemma1}.
\end{lem}
\begin{lem}
	\cite{chen2015completing} Given the assumptions in Lemma \ref{Lemma1}, for any matrix $\mathbf {Z} \in \mathbf {T}$, we have the following inequality with high probability:
	$$ \|(\mathcal{P}_{\mathbf{T}} - \mathcal{P}_{\mathbf{T}}\circ \mathcal{R} _\Omega\circ\mathcal{P}_{\mathbf{T}})(\mathbf{Z}) \|_{\mu(\infty)} \le \tfrac{1}{2}\| \mathbf{Z} \|_{\mu(\infty)} $$
	In some cases, we will use $\overline{\mathcal{R}}_\Omega$ instead of $\mathcal{R} _\Omega$ which is defined as:
	\begin{align}
	\overline{\mathcal{R}}_\Omega(\overline{\mathbf{Z}}) \triangleq \mathbf{B}_L^{\rm H} \mathcal{R} _\Omega( \mathbf{B}_L\overline{\mathbf{Z}} \mathbf{B}_R^{\rm H}) \mathbf{B}_R \label{A40}
	\end{align}
	Based on the sampling operator $\mathcal{R} _\Omega$, we establish the orthogonal projection $\mathcal{P}_{\Omega}(\mathbf {Z}) $ in order to project the input to the support of $\mathcal{R} _\Omega$:
	\begin{align}
 	&\mathcal{P}_{\Omega}(\mathbf{Z})  \triangleq\sum_{i,l=1}^{n}\epsilon_{il}\mathbf{Z}[i,l]\mathbf{e}_{i}\mathbf{e}_l^{T}.
	\end{align} 
	Similar to the definition of  $\overline{\mathcal{R}}_\Omega$ and $\mathcal{P}_{\Omega}$, we have:
	\begin{align}
	&\overline{\mathcal{P}}_{\Omega}(\overline{\mathbf{Z}}) = \mathbf{B}_L^{\rm H} \mathcal{P}_{\Omega}( \mathbf{B}_L\overline{\mathbf{Z}} \mathbf{B}_R^{\rm H}) \mathbf{B}_R. \label{A42} 
	\end{align}
\end{lem}
In the following, we provide some main features of the recently defined variables.
\begin{lem}\cite{eftekhari2018weighted}
	For an arbitrary matrix $\mathbf {Z}$ and $\overline{\mx{Z}}$, 
 we have:
	\begin{align}
	\langle \overline{\mathbf{Z}},\overline{\mathcal{R}}_\Omega(\overline{\mathbf{Z}})\rangle = \langle \mathbf{Z},\mathcal{R} _\Omega(\mathbf{Z})\rangle
	\end{align}
	\begin{align}
	\| \overline{\mathcal{R}}_\Omega(\overline{\mathbf{Z}})\|_{F} = \| \mathcal{R} _\Omega(\mathbf{Z})\|_{F}
	\end{align} 
	Also if $0 \leq k \leq h \leq1$ and $p_{i,l}\subset [l,h]$, then
	\begin{align}
	(\overline{\mathcal{R}}_\Omega\circ\overline{\mathcal{R}}_\Omega)(\cdot) \succeq (\overline{\mathcal{R}}_\Omega)(\cdot)
	\end{align} 
	\begin{align}
	\| \overline{\mathcal{R}}_\Omega(\cdot) \|_{F \rightarrow F} =\| \mathcal{R} _\Omega(\cdot) \|_{F \rightarrow F} \le l^{-1} 
	\end{align} 
	where $\mathcal{A}(\cdot) \succeq \mathcal{B}(\cdot)$ means that for every matrix $\mathbf {Z}$, $\langle \mathbf {Z} , \mathcal A (\mathbf {Z}) \rangle \ge \langle \mathbf {Z }, \mathcal B (\mathbf {Z}) \rangle $.

\end{lem}
The latter lemma results in the following relation:
	\begin{align}
	\|\overline{\mathcal{P}}_{\Omega}(\overline{\mathbf{Z}})\|_{F} \le h\|\overline{\mathcal{R}}_\Omega(\overline{\mathbf{Z}})\|_{F}.
	\end{align}
Now, with these lemmas in mind, we can finalize the proof of Theorem \eqref{theory_main}.

Consider $\mathbf {X} = \mathbf{X}_r$ is a rank-$r$ matrix and our observation is measured in a noisy environment. Let $\hat{\mathbf X}$ and $\bs{\Psi}\triangleq \hat{\mathbf {X}} - \mathbf {X}$ be the solution and the estimation error of \eqref{prior_nuclear_prob}. Then one can write:
\begin{align}
\| \mathbf{Q}_{\widetilde{\bm{\mathcal{U}}}_{r}} (\mathbf{X}+\bs{\Psi}) \mathbf{Q}_{\widetilde{\bm{\mathcal{V}}}_{r}} \|_{*} \le  
\| \mathbf{Q}_{\widetilde{\bm{\mathcal{U}}}_{r}} \mathbf{X} \mathbf{Q}_{\widetilde{\bm{\mathcal{V}}}_{r}} \|_{*} \label{A48}
\end{align}
for the right side of \eqref{A48} we have:
\begin{align}
&\| \mathbf{Q}_{\widetilde{\bm{\mathcal{U}}}_{r}} \mathbf{X} \mathbf{Q}_{\widetilde{\bm{\mathcal{V}}}_{r}} \|_{*} = \| \mathbf{Q}_{\widetilde{\bm{\mathcal{U}}}_{r}} \mathbf{X}_r \mathbf{Q}_{\widetilde{\bm{\mathcal{V}}}_{r}} \|_{*}  \nonumber \\
& \le \| \mathbf{B}_L \mathbf{O}_L \mathbf{L}  \overline{\mathbf{X}}_{r} \mathbf{R}^{\rm H} \mathbf{O}_R^{\rm H}  \mathbf{B}_R^{\rm H} \|_{*} =  \| \mathbf{L}  \overline{\mathbf{X}}_{r} \mathbf{R}^{\rm H} \|_{*}  \nonumber \\
&= \Bigg\|
\begin{bmatrix}
\mathbf{L}_{11}\overline{\mathbf{X}}_{r,11}\mathbf{R}_{11}  & \\
& \mathbf{0}_{n-r}
\end{bmatrix}
\Bigg\|_{*}
\end{align}
which holds since $\mathbf{B}_R, \mathbf{O}_R, \mathbf{B}_L$ and $\mathbf{O}_L$ are orthogonal. For the left-hand side of \eqref{A48}, we have:
\begin{align}
&\| \mathbf{Q}_{\widetilde{\bm{\mathcal{U}}}_{r}} (\mathbf{X}+\bs{\Psi}) \mathbf{Q}_{\widetilde{\bm{\mathcal{V}}}_{r}} \|_{*} = \| \mathbf{Q}_{\widetilde{\bm{\mathcal{U}}}_{r}} (\mathbf{X}_r+\bs{\Psi}) \mathbf{Q}_{\widetilde{\bm{\mathcal{V}}}_{r}} \|_{*}  \nonumber \\
&\ge \|   \mathbf{B}_L \mathbf{O}_L \mathbf{L} (\overline{\mathbf{X}}_{r} + \overline{\bs{\Psi}}) \mathbf{R}^{\rm H} \mathbf{O}_R^{\rm H}  \mathbf{B}_R^{\rm H} \|_{*} - \|\mathbf{X}_{r}^{+} \|_{*}  \nonumber \\
&= \|  \mathbf{L} (\overline{\mathbf{X}}_{r} + \overline{\bs{\Psi}}) \mathbf{R}^{\rm H} \|_{*} - \|\mathbf{X}_{r}^{+} \|_{*}  \nonumber \\
&= \Bigg\|
\begin{bmatrix}
\mathbf{L}_{11}\overline{\mathbf{X}}_{r,11}\mathbf{R}_{11}  & \\
& \mathbf{0}_{n-r} \end{bmatrix} + \mathbf{L}\overline{\bs{\Psi}} \mathbf{R}^{\rm H}
\Bigg\|_{*} - \|\mathbf{X}_{r}^{+} \|_{*}
 \end{align}
Replacing these two upper and lower bounds in \eqref{A48} and considering the convexity of nuclear norm, we have:
\begin{align}
&\langle \mathbf{L} \overline{\bs{\Psi}} \mathbf{R}^{\rm H}, \mathbf{G}^{\rm H} \rangle  \nonumber \\
&\le \Bigg\|
\begin{bmatrix}
\mathbf{L}_{11}\overline{\mathbf{X}}_{r,11}\mathbf{R}_{11}  & \\
& \mathbf{0}_{n-r} \end{bmatrix} + \mx{L}\overline{\bs{\Psi}} \mathbf{R}^{\rm H}
\Bigg\|_{*}  \nonumber \\
& - \Bigg\|
\begin{bmatrix}
\mathbf{L}_{11}\overline{\mathbf{X}}_{r,11}\mathbf{R}_{11}  & \\
 & \mathbf{0}_{n-r} \end{bmatrix}
\Bigg\|_{*} \le 0  \nonumber \\
 & \forall \mathbf{G} \in \partial\Bigg\|
 \begin{bmatrix}
 \mathbf{L}_{11}\overline{\mathbf{X}}_{r,11}\mathbf{R}_{11}  & \\
 & \mathbf{0}_{n-r} \end{bmatrix}
 \Bigg\|_{*} \label{A51}
\end{align}
In \eqref{A51}, $ \partial\| \mathbf {A }\|_*$ indicates the sub-differential of nuclear norm at point $\mathbf{ A}$ \cite{recht2010guaranteed}. In order to fully describe sub-differential, let ${\rm rank}(\overline{\mathbf{X}}_{r,11})={\rm rank}({\mathbf {X}}_r) = r$ and for non-zero weights ${\rm rank}(\mathbf{L}_{11} \overline{\mathbf {X}}_{11} \mathbf{R}_{11})=r$. Consider the SVD decomposition for matrix $\mathbf{L}_{11} \overline{\mathbf X}_{r,11} \mathbf{R}_{11}$ as:
$$ \mathbf{L}_{11}\overline{\mathbf{X}}_{r,11}\mathbf{R}_{11} = \overline{\mathbf{U}}_r \overline{\bs{\Delta}}_r \overline{\mathbf{V}}_r^{\rm H} $$
Also let $\mathbf S$ be the sign matrix defined as:
\begin{align}
\mathbf{S}\triangleq \begin{bmatrix}
\mathbf{S}_{11} & \\
& \mathbf{0}_{n-r}
\end{bmatrix} \triangleq \begin{bmatrix}
\overline{\mathbf{U}}_r \overline{\mathbf{V}}_r^{\rm H} & \\
& \mathbf{0}_{n-r}
\end{bmatrix} \label{S-matrix}
\end{align} 
Finally, the sub-differential is defined as:
\begin{align}
&\partial\left\|
\begin{bmatrix}
\mathbf{L}_{11}\overline{\mathbf{X}}_{r,11}\mathbf{R}_{11}  & \\
& \mathbf{0}_{n-r} \end{bmatrix}
\right\|_{*}  \nonumber \\
&= \Bigg\{\mathbf{G} \in \mathbb{R}^{n \times n }:\nonumber\\
&\mathbf{G} = \begin{bmatrix}
	\mathbf{S}_{11} \in \mathbb{R}^{n \times n}&\\
	& \mathbf{G}_{22} \in \mathbb{R}^{(n-r) \times (n-r)}
	\end{bmatrix},  \|\mathbf{G}_{22}\| \le 1\Bigg\}
 =  \nonumber \\
&\Bigg\{ \mathbf{G} \in \mathbb{R}^{n \times n } : \mathcal{P}_{\overline{\mathbf{T}}}(\mathbf{G}) = \mathbf{S} =  \begin{bmatrix}
\mathbf{S}_{11} & \\
& \mathbf{0}_{n-r}
\end{bmatrix}, \mathcal{P}_{\overline{\mathbf{T}}^{\perp}}(\mathbf{G}) \le 1\Bigg\} .
\end{align}
By considering \eqref{S-matrix}, we have:
\begin{align}
&{\rm rank}(\mathbf{S}) = {\rm rank}(\mathbf{S}_{11}) =r, \|\mathbf{S}\| = \|\mathbf{S}_{11}\| = 1 \nonumber \\
&\|\mathbf{S}\|_{F} = \|\mathbf{S}_{11}\|_{F} = \sqrt{r}
\end{align}
Then according to the aforementioned equations and for $\|\mathbf{G}_{22} \| \leq 1$, \eqref{A51} becomes as:
\begin{align}
\langle \mathbf{L}\overline{\bs{\Psi}}\mathbf{R}^{\rm H} , \mathbf{S}+ & \mathbf{0}_{n-r} \rangle \le 0 
\end{align}
which is simplified to
\begin{align}
&0 \ge  \langle \mathbf{L}\overline{\bs{\Psi}}\mathbf{R}^{\rm H} , \mathbf{S}  \rangle + \sup_{\|\mathbf{G}_{22}\| \le 1} \langle \mathbf{L}\overline{\bs{\Psi}}\mathbf{R}^{\rm H} ,  \begin{bmatrix}
\mathbf{0}_{r} & \\
& \mathbf{G}_{22}
\end{bmatrix} \rangle  \nonumber \\
&= \langle \mathbf{L}\overline{\bs{\Psi}}\mathbf{R}^{\rm H} , \mathbf{S}  \rangle + \sup_{\|\mathbf{G}\| \le 1} \langle \mathcal{P}_{\overline{\mathbf{T}}^{\perp}}(\mathbf{L}\overline{\bs{\Psi}}\mathbf{R}^{\rm H}) ,  \mathbf{G} \rangle  \nonumber \\
&= \langle \mathbf{L}\overline{\bs{\Psi}}\mathbf{R}^{\rm H} , \mathbf{S}  \rangle + \| \mathcal{P}_{\overline{\mathbf{T}}^{\perp}}(\mathbf{L}\overline{\bs{\Psi}}\mathbf{R}^{\rm H}) \|_{*}   \nonumber \\
&= \langle \overline{\bs{\Psi}} , \mathbf{L}^{\rm H}\mathbf{S}\mathbf{R}\rangle + \| \mathcal{P}_{\overline{\mathbf{T}}^{\perp}}(\mathbf{L}\overline{\bs{\Psi}}\mathbf{R}^{\rm H}) \|_{*}  \nonumber \\
&=\Bigg\langle \overline{\bs{\Psi}} , \begin{bmatrix}
\mathbf{L}_{11}\mathbf{S}_{11}\mathbf{R}_{11} & \mathbf{L}_{11}\mathbf{S}_{11}\mathbf{R}_{12} & \\
\mathbf{L}_{12}^{\rm H}\mathbf{S}_{11}\mathbf{R}_{11} & \mathbf{L}_{12}^{\rm H}\mathbf{S}_{11}\mathbf{R}_{12}& \\
& & \mathbf{0}_{n-2r}
\end{bmatrix}\Bigg\rangle  \nonumber \\
& +\Bigg\|
\begin{bmatrix}
\mathbf{0}_{r}&&&\\
 & \mathbf{L}_{22}\overline{\bs{\Psi}}_{22}\mathbf{R}_{22} & \mathbf{L}_{22}\overline{\bs{\Psi}}_{23}\bs{\Gamma}_2 & \mathbf{L}_{22}\overline{\bs{\Psi}}_{24} \\
&\bs{\Lambda}_2\overline{\bs{\Psi}}_{32}\mathbf{R}_{22}& \bs{\Lambda}_2\overline{\bs{\Psi}}_{33}\bs{\Gamma}_2 & \bs{\Lambda}_2\overline{\bs{\Psi}}_{34}\\
&\overline{\bs{\Psi}}_{42}\mathbf{R}_{22}& \overline{\bs{\Psi}}_{34}\bs{\Gamma}_2 & \overline{\bs{\Psi}}_{44}
\end{bmatrix}
\Bigg\|_{*}  \nonumber \\
& =\Bigg\langle \overline{\bs{\Psi}} , \begin{bmatrix}
\mathbf{L}_{11}\mathbf{S}_{11}\mathbf{R}_{11} & \mathbf{L}_{11}\mathbf{S}_{11}\mathbf{R}_{12} & \\
\mathbf{L}_{12}^{\rm H}\mathbf{S}_{11}\mathbf{R}_{11}& \mathbf{0}_r \\
& & \mathbf{0}_{n-2r}
\end{bmatrix}\Bigg\rangle  \nonumber \\
& + \langle \overline{\bs{\Psi}}_{22}, \mathbf{L}_{12}^{\rm H}\mathbf{S}_{11}\mathbf{R}_{12} \rangle  \nonumber \\
& \Bigg\|
\begin{bmatrix}
\mathbf{0}_{r}&&& \\
 & \mathbf{L}_{22}\overline{\bs{\Psi}}_{22}\mathbf{R}_{22}& \mathbf{L}_{22}\overline{\bs{\Psi}}_{23}\bs{\Gamma}_2 & \mathbf{L}_{22}\overline{\bs{\Psi}}_{24} \\
 &\bs{\Lambda}_2\overline{\bs{\Psi}}_{32}\mathbf{R}_{22}& \bs{\Lambda}_2\overline{\bs{\Psi}}_{33}\bs{\Gamma}_2 & \bs{\Lambda}_2\overline{\bs{\Psi}}_{34}\\
&\overline{\bs{\Psi}}_{42}\mathbf{R}_{22}& \overline{\bs{\Psi}}_{34}\bs{\Gamma}_2 & \overline{\bs{\Psi}}_{44}
\end{bmatrix}
\Bigg\|_{*}  \nonumber \\
&=:\langle \overline{\bs{\Psi}} , \overline{\mathbf{S}}^{\prime}\rangle + \langle\overline{\bs{\Psi}}_{22}, \mathbf{L}_{12}^{\rm H}\mathbf{S}_{11}\mathbf{R}_{12} \rangle + \| \mathbf{L}^{\prime}\mathcal{P}_{\overline{\mathbf{T}}^{\perp}}(\overline{\bs{\Psi}}) \mathbf{R}^{\prime} \|_{*} \label{A56}
\end{align} 
where \eqref{A56} holds due to the fact that the spectral norm is the dual function of nuclear norm. Also the matrices $\mathbf{S}'$ and $\mathbf{L}'$ are defined as below:
$$ \overline{\mathbf{S}}^{\prime} \triangleq \begin{bmatrix}
\mathbf{L}_{11}\mathbf{S}_{11}\mathbf{R}_{11} & \mathbf{L}_{11}\mathbf{S}_{11}\mathbf{R}_{12} & \\
\mathbf{L}_{12}^{\rm H}\mathbf{S}_{11}\mathbf{R}_{11} & \mathbf{0}_r & \\
& & \mathbf{0}_{n-r^{\prime}-r}
\end{bmatrix} $$
$$\mathbf{L}^{\prime} \triangleq \begin{bmatrix}
\mathbf{0}_{r} &  & \\
& \mathbf{L}_{22} & & \\
 & & \bs{\Lambda}_2& \\
& & & \mathbf{I}_{n-r^{\prime}	-2r}
\end{bmatrix} $$
It is worth mentioning that $\overline{\mx{S}}\in\overline{\mathbf T}$. Regarding $\overline{\mathbf S}'$, we may write:
\begin{align}
&0 \ge \langle \overline{\bs{\Psi}} , \overline{\mathbf{S}}^{\prime} \rangle + \langle \overline{\bs{\Psi}}_{22} , \mathbf{L}_{12}^{\rm H}\mathbf{S}_{11}\mathbf{R}_{12} \rangle + \| \mathbf{L}^{\prime}\mathcal{P}_{\overline{\mathbf{T}}^{\perp}}(\overline{\bs{\Psi}}) \mathbf{R}^{\prime} \|_{*}  \nonumber \\
& \ge \langle \overline{\bs{\Psi}} , \overline{\mathbf{S}}^{\prime} \rangle  - \|\mathcal{P}_{\overline{\mathbf{T}}^{\perp}}(\overline{\bs{\Psi}}) \|_{*} \|\mathbf{L}_{12}\| \|\mathbf{S}_{11}\| \|\mathbf{R}_{12}\| + \|\mathcal{P}_{\overline{\mathbf{T}}^{\perp}}(\overline{\bs{\Psi}}) \|_{*}  \nonumber \\
& - 
\| \mathcal{P}_{\overline{\mathbf{T}}^{\perp}}(\overline{\bs{\Psi}}) -  \mathbf{L}^{\prime}\mathcal{P}_{\overline{\mathbf{T}}^{\perp}}(\overline{\bs{\Psi}}) \mathbf{R}^{\prime} \|_{*}  \nonumber \\
& = \langle \overline{\bs{\Psi}} , \overline{\mathbf{S}}^{\prime} \rangle + (1-\|\mathbf{L}_{12}\|\|\mathbf{R}_{12}\|)\|\mathcal{P}_{\overline{\mathbf{T}}^{\perp}}(\overline{\bs{\Psi}}) \|_{*}  \nonumber \\
&- \|\mathcal{P}_{\overline{\mathbf{T}}^{\perp}}(\mathbf{I}_{n}) \mathcal{P}_{\overline{\mathbf{T}}^{\perp}} 
(\overline{\bs{\Psi}})\mathcal{P}_{\overline{\mathbf{T}}^{\perp}}(\mathbf{I}_{n}) -  \mathbf{L}^{\prime}\mathcal{P}_{\overline{\mathbf{T}}^{\perp}}(\overline{\bs{\Psi}}) \mathbf{R}^{\prime}\|_{*}
\nonumber\\ & \ge \langle \overline{\bs{\Psi}} , \overline{\mathbf{S}}^{\prime} \rangle + (1-\|\mathbf{L}_{12}\|\|\mathbf{R}_{12}\|)\|\mathcal{P}_{\overline{\mathbf{T}}^{\perp}}(\overline{\bs{\Psi}}) \|_{*} \nonumber \\
& -  \|\mathcal{P}_{\overline{\mathbf{T}}^{\perp}}(\mathbf{I}_{n}) - \mathbf{L}^{\prime} \| \|\mathcal{P}_{\overline{\mathbf{T}}^{\perp}}(\overline{\bs{\Psi}}) \|_{*}
\nonumber \\
& - \| \mathbf{L}^{\prime} \| \|\mathcal{P}_{\overline{\mathbf{T}}^{\perp}}(\overline{\bs{\Psi}}) \|_{*} \|\mathcal{P}_{\overline{\mathbf{T}}^{\perp}}(\mathbf{I}_{n}) - \mathbf{R}^{\prime} \| \nonumber \\
& \ge \langle \overline{\bs{\Psi}} , \overline{\mathbf{S}}^{\prime} \rangle + (1-\|\mathbf{L}_{12}\|\|\mathbf{R}_{12}\|)\|\mathcal{P}_{\overline{\mathbf{T}}^{\perp}}(\overline{\bs{\Psi}}) \|_{*} \nonumber \\
& -  \Big\|\begin{bmatrix}
\mathbf{I}_{r} - \mathbf{L}_{22} & \\
& \mathbf{I}_{r} - \bs{\Lambda}_{2}
\end{bmatrix} \Big\| \|\mathcal{P}_{\overline{\mathbf{T}}^{\perp}}(\overline{\bs{\Psi}}) \|_{*}
- \|\mathcal{P}_{\overline{\mathbf{T}}^{\perp}}(\overline{\bs{\Psi}}) \|_{*} \nonumber \\
& \quad \quad \quad \quad \quad \quad \quad \quad \quad \Big\|\begin{bmatrix}
\mathbf{I}_{r} - \mathbf{R}_{22} & \\
& \mathbf{I}_{r} - \bs{\Gamma}_{2}
\end{bmatrix} \Big\| \nonumber \\
& = \langle \overline{\bs{\Psi}} , \overline{\mathbf{S}}^{\prime} \rangle  + (1-\|\mathbf{L}_{12}\|\|\mathbf{R}_{12}\|-  \Big\|\begin{bmatrix}
\mathbf{I}_{r} - \mathbf{L}_{22} & \\
& \mathbf{I}_{r} - \bs{\Lambda}_{2}
\end{bmatrix} \Big\| \nonumber \\
&\quad \quad \quad \quad \quad \quad \quad -\Big\|\begin{bmatrix}
\mathbf{I}_{r} - \mathbf{R}_{22} & \\
& \mathbf{I}_{r} - \bs{\Gamma}_{2} \end{bmatrix} \Big\|) \|\mathcal{P}_{\overline{\mathbf{T}}^{\perp}}(\overline{\bs{\Psi}}) \|_{*} \nonumber \\
&  \stackrel{\eqref{def_fun}}{=}\langle \overline{\bs{\Psi}} , \overline{\mathbf{S}}^{\prime} \rangle + \|\mathcal{P}_{\overline{\mathbf{T}}^{\perp}}(\overline{\bs{\Psi}}) \|_{*} \nonumber \\
&\Bigg[1- \sqrt{\max\Big\{\tfrac{f^2_2(1-\lambda^2_{1}(i) , \theta_u(i))}{f^2_2(\lambda_{1}(i) , \theta_u(i))}\Big\}} \nonumber \\
&\quad \quad \quad \quad \quad  \sqrt{\max\Big\{\tfrac{f^2_2(1-\gamma_1^2(i) , \theta_v(i))}{f^2_2(\gamma_1(i) , \theta_v(i))}\Big\}} \nonumber \\
&-\max_i \Big\{  \max_i \{\lambda_{2}(i)-1\} , \max_i \Big\{\tfrac{\lambda_{1}(i)}{f_2(\lambda_1(i) , \theta_u(i))}-1 \Big\}  \Big\} \nonumber\\
& -\max_i \Big\{  \max_i \{\gamma_{2}(i)-1\}  , \max_i \Big\{\tfrac{\gamma_{1}(i)}{f_2(\gamma_1(i) , \theta_v(i))}-1 \Big\} \Big\}\Bigg]\nonumber\\ 
& =: \langle \overline{\bs{\Psi}} , \overline{\mathbf{S}}^{\prime} \rangle + (1-\alpha_3(\theta_{u}(i),\theta_{v}(i),\lambda_{i},\gamma_i))\|\mathcal{P}_{\overline{\mathbf{T}}^{\perp}}(\overline{\bs{\Psi}}) \|_{*} \label{A58}
\end{align}
in which the first inequality holds due to \eqref{A56}, the second one comes from holder's inequality, and the fact that $\overline{\mathbf H}_{22}$ is a part of $\mathcal{P}_{\overline{T}^{\perp}}(\overline{\mathbf H})$. The last inequality is due to $ab+a+b \leq \tfrac{3}{2}(a+b)$. 

The following lemma defines the dual certificate.
\begin{lem} \label{lemmaA-2-6}
Let $\overline{\mathbf T}$ be the support of matrix $\overline{\mathbf X}_r$ as defined in \eqref{3-42} and
$\underset{i,l}{\rm min}~ p_{il} \leq k$. Consider $\overline{\mathcal{R}}_\Omega$ and $\overline{\mathcal{P}}_{\Omega}$ as defined in \eqref{A40} and \eqref{A42}, respectively. Then, for $i,l \in [1:n]$, as long as
\begin{align}
 &\max [\log(\alpha_1n),1]\cdot \tfrac{(\mu_i+\nu_l)r\log n}{n} \nonumber \\ 
&\cdot \max[\alpha_2 \Big(1+ \max_{i} \tfrac{\breve{\mu}_i}{\mu_i} + \max_{l} \tfrac{\breve{\nu}_l }{\nu_l}\Big) ,1] \lesssim p_{il} \le 1,
\end{align}
then, we have:
\begin{align} \|(\mathcal{P}_{\overline{\mathbf{T}}} - \mathcal{P}_{\overline{\mathbf{T}}}\circ \overline{\mathcal{R}}_{\Omega}\circ\mathcal{P}_{\overline{\mathbf{T}}})(\cdot) \|_{F \rightarrow F} \le \tfrac{1}{2},\label{A60}
\end{align}
and there exists a matrix $\overline{\bs{\Pi}} \in \mathbb{R}^{n \times n}$ that satisfies the following relations:
\begin{align}
&\| \overline{\mathbf{S}}^{\prime} - \mathcal{P}_{\overline{\mathbf{T}}}(\overline{\bs{\Pi}}) \|_{F} \le \tfrac{k}{4\sqrt{2}} \label{A61}\\
&\mathcal{P}_{\overline{\mathbf{T}}^{\perp}}(\overline{\bs{\Pi}}) \le \tfrac{1}{2} \label{A62} \\ & \overline{\bs{\Pi}} = \overline{\mathcal{R}}_{\Omega}(\overline{\bs{\Pi}}), \label{A63}
\end{align}
where $\alpha_1$ and $\alpha_2$ are defined as:
\begin{align}
&\alpha_1 = \alpha_1(\theta_{u}(i),\theta_{v}(i),\lambda_{1}(i),\gamma_{1}(i))
\triangleq\nonumber \\
& \sqrt{\underset{i}{\rm max}\bigg\{\tfrac{f^2_1(\lambda_1(i) , \theta_u(i))} {f^2_2(\lambda_1(i) , \theta_u(i))} \bigg\} \cdot \underset{i}{\rm max}\bigg\{\tfrac{f^2_1(\gamma_1(i) , \theta_v(i))} {f^2_2(\gamma_1(i) , \theta_v(i))}\bigg\}} \nonumber \\
&\alpha_2 = \alpha_2(\theta_{u}(i),\theta_{v}(i),\lambda_{1}(i),\gamma_{1}(i))
\triangleq \nonumber \\
&\sqrt{\underset{i}{\rm max} \Big\{f^2_2(\lambda_1(i) , \theta_u(i)\Big\}\underset{i}\cdot{\rm max} \bigg\{\tfrac{f^2_1(\gamma_1(i) , \theta_v(i))} {f^2_2(\gamma_1(i) , \theta_v(i))}\bigg\}}\nonumber \\
&~~~~~~~~~~~~~~~~~~~~~~~~ \cdot \Big( 1 + \sqrt{\tfrac{r^{\prime}+r}{r} \max_{i}\tfrac{\breve{\mu}_i}{\mu_i}}\Big) \nonumber \\
& +  \sqrt{\underset{i}{\rm max}\Big\{f^2_2(\gamma_1(i) , \theta_v(i))\Big\} \underset{i}\cdot {\rm max} \bigg\{\tfrac{f^2_1(\lambda_1(i) , \theta_u(i))} {f^2_2(\lambda_1(i) , \theta_u(i))}\bigg\}}
\nonumber \\
&~~~~~~~~~~~~~~~~~~~~~\cdot \Big( 1+ \sqrt{\tfrac{r^{\prime}+r}{r} \max_{i}\tfrac{\breve{\nu}_i}{\nu_i}}\Big) \nonumber \\
\blacksquare
\end{align}
\end{lem}
According to Lemma \ref{lemmaA-2-6}, there exists a $\overline{\bs{\Pi}}$ satisfying \eqref{A61} to \eqref{A63}. Thus, \eqref{A56} can be rewritten as:
\begin{align}
&0 \ge  \langle \overline{\bs{\Psi}} , \overline{\mathbf{S}}^{\prime} \rangle + (1-\alpha_3)\|\mathcal{P}_{\overline{\mathbf{T}}^{\perp}}(\overline{\bs{\Psi}}) \|_{*} \nonumber \\
& = \langle \overline{\bs{\Psi}} , \mathcal{P}_{\overline{\mathbf{T}}}(\overline{\bs{\Pi}}) \rangle + \langle \overline{\bs{\Psi}} , \overline{\mathbf{S}}^{\prime} -\mathcal{P}_{\overline{\mathbf{T}}}(\overline{\bs{\Pi}}) \rangle + (1-\alpha_3)\|\mathcal{P}_{\overline{\mathbf{T}}^{\perp}}(\overline{\bs{\Psi}}) \|_{*} \nonumber \\ 
& = \langle \overline{\bs{\Psi}} , \overline{\bs{\Pi}} \rangle +  \langle \overline{\bs{\Psi}} , \mathcal{P}_{\overline{\mathbf{T}}}(\overline{\bs{\Pi}}) \rangle + \langle \overline{\bs{\Psi}} , \overline{\mathbf{S}}^{\prime} -\mathcal{P}_{\overline{\mathbf{T}}}(\overline{\bs{\Pi}}) \rangle+ \nonumber \\
&(1-\alpha_3)\|\mathcal{P}_{\overline{\mathbf{T}}^{\perp}}(\overline{\bs{\Psi}}) \|_{*}  = -  \langle \overline{\bs{\Psi}} , \mathcal{P}_{\overline{\mathbf{T}}^{\perp}}(\overline{\bs{\Pi}}) \rangle + \langle \overline{\bs{\Psi}} , \overline{\mathbf{S}}^{\prime} -\mathcal{P}_{\overline{\mathbf{T}}}(\overline{\bs{\Pi}}) \rangle 
\nonumber \\
&+ (1-\alpha_3)\|\mathcal{P}_{\overline{\mathbf{T}}^{\perp}}(\overline{\bs{\Psi}}) \|_{*}  \ge \|\mathcal{P}_{\overline{\mathbf{T}}^{\perp}}(\overline{\bs{\Psi}}) \|_{*} \|\mathcal{P}_{\overline{\mathbf{T}}^{\perp}}(\overline{\bs{\Pi}})\|\nonumber \\ &-\|\mathcal{P}_{\overline{\mathbf{T}}}(\overline{\bs{\Psi}})\|_{f}\| \overline{\mathbf{S}}^{\prime} -\mathcal{P}_{\overline{\mathbf{T}}}(\overline{\bs{\Pi}})  \|_{F}
+ (1-\alpha_3)\|\mathcal{P}_{\overline{\mathbf{T}}^{\perp}}(\overline{\bs{\Psi}})\|_*
\nonumber \\
& \ge\tfrac{-1}{2}\|\mathcal{P}_{\overline{\mathbf{T}}^{\perp}}(\overline{\bs{\Psi}})\|_{*} -\tfrac{k}{4\sqrt{2}}\|\mathcal{P}_{\overline{\mathbf{T}}}(\overline{\bs{\Psi}})\|_{F}+(1-\alpha_3)\|\mathcal{P}_{\overline{\mathbf{T}}^{\perp}}(\overline{\bs{\Psi}})\|_{*}\nonumber \\
& = (\tfrac{1}{2} - \alpha_3)\|\mathcal{P}_{\overline{\mathbf{T}}^{\perp}}(\overline{\bs{\Psi}})\|_{*}- \tfrac{k}{4\sqrt{2}}\|\mathcal{P}_{\overline{\mathbf{T}}}(\overline{\bs{\Psi}})\|_{F}.\label{A65}
\end{align}
For $\alpha_3 \leq 1$, \eqref{A65} is equivalent to:
\begin{align}
(\tfrac{1}{2} - \alpha_3)\|\mathcal{P}_{\overline{\mathbf{T}}^{\perp}}(\overline{\bs{\Psi}})\|_{*} \le \tfrac{k}{4\sqrt{2}} \|\mathcal{P}_{\overline{\mathbf{T}}}(\overline{\bs{\Psi}})\|_{F}. \label{A66}
\end{align}
Here, the third inequality of \eqref{A65} comes from Holder's inequality, the equations \eqref{A61} and \eqref{A62} 
 and the following relations:
\begin{align}
\| \overline{\mathcal{R}}_p(\overline{\bs{\Psi}}) \|_{F} = \| \mathcal{R} _\Omega(\bs{\Psi}) \|_{F} = \| \mathcal{R} _\Omega(\widehat{\mathbf{X}} - \mathbf{X}) \|_{F} = 0 ,
\end{align}
\begin{align}
\langle \overline{\bs{\Psi}} , \overline{\bs{\Pi}} \rangle = \langle \overline{\bs{\Psi}} , \overline{\mathcal{P}}_{\Omega}(\overline{\bs{\Pi}}) \rangle = \langle \overline{\mathcal{P}}_{\Omega}(\overline{\bs{\Psi}}) , \overline{\bs{\Pi}} \rangle  =  0 .
\end{align} 
In order to calculate the upper-bound of \eqref{A66}, it holds that:
\begin{align}
&\| \overline{\mathcal{R}}_{\Omega}(\mathcal{P}_{\overline{\mathbf{T}}}(\overline{\bs{\Psi}})) \|_{F} = \| \overline{\mathcal{R}}_{\Omega}(\mathcal{P}_{\overline{\mathbf{T}}^{\perp}}(\overline{\bs{\Psi}})) \|_{F}   \nonumber \\
&\le 
\| \overline{\mathcal{R}}_{\Omega}(\cdot) \|_{F \rightarrow F} \| \mathcal{P}_{\overline{\mathbf{T}}^{\perp}}({\bs{\Psi}}) \| _{F}
\le \tfrac{1}{k} \| \mathcal{P}_{\overline{\mathbf{T}}^{\perp}}(\overline{\bs{\Psi}}) \| _{F}\label{A69}
\end{align}
Now by considering Lemma \ref{lemmaA-2-6}, it follows that
\begin{align}
&\| \overline{\mathcal{R}}_{\Omega}(\mathcal{P}_{\overline{\mathbf{T}}}(\overline{\bs{\Psi}})) \|_{F}^{2} = \langle \mathcal{P}_{\overline{\mathbf{T}}}(\overline{\bs{\Psi}}) ,(\overline{\mathcal{R}}_{\Omega} \circ \overline{\mathcal{R}}_{\Omega}	)(\mathcal{P}_{\overline{\mathbf{T}}}(\overline{\bs{\Psi}})) \rangle  \nonumber \\
&\ge \langle \mathcal{P}_{\overline{\mathbf{T}}}(\overline{\bs{\Psi}}) , \overline{\mathcal{R}}_{\Omega}	(\mathcal{P}_{\overline{\mathbf{T}}}(\overline{\bs{\Psi}})) \rangle = \langle \mathcal{P}_{\overline{\mathbf{T}}}(\overline{\bs{\Psi}}) , \mathcal{P}_{\overline{\mathbf{T}}}(\overline{\bs{\Psi}}) \rangle  \nonumber \\
& + \langle \mathcal{P}_{\overline{\mathbf{T}}}(\overline{\bs{\Psi}}) , ( \mathcal{P}_{\overline{\mathbf{T}}} \circ  \overline{\mathcal{R}}_{\Omega} \circ \mathcal{P}_{\overline{\mathbf{T}}} - \mathcal{P}_{\overline{\mathbf{T}}}) \circ \mathcal{P}_{\overline{\mathbf{T}}}(\overline{\bs{\Psi}}) \rangle \nonumber \\
& \ge \| \mathcal{P}_{\overline{\mathbf{T}}}(\overline{\bs{\Psi}}) \|_{F}^{2} - \| \mathcal{P}_{\overline{\mathbf{T}}} \circ  \overline{\mathcal{R}}_p \circ \mathcal{P}_{\overline{\mathbf{T}}} - \mathcal{P}_{\overline{\mathbf{T}}} \|_{F \rightarrow F} \| \mathcal{P}_{\overline{\mathbf{T}}}(\overline{\bs{\Psi}})\|_{F}^{2}
\nonumber \\
&\ge \tfrac{1}{2} \| \mathcal{P}_{\overline{\mathbf{T}}}(\overline{\bs{\Psi}})\|_{F}^{2} \label{A70}
\end{align} 
Thus, combining \eqref{A69} and \eqref{A70} leads to:
\begin{align}
\| \mathcal{P}_{\overline{\mathbf{T}}}(\overline{\bs{\Psi}})\|_{F} \le \tfrac{\sqrt{2}}{k} \| \mathcal{P}_{\overline{\mathbf{T}}^{\perp}}(\overline{\bs{\Psi}})\|_{F}. \label{A71}
\end{align}
Finally, by considering \eqref{A71} and \eqref{A66}, we have:
\begin{align}
(\tfrac{1}{2} - \alpha_3) \| \mathcal{P}_{\overline{\mathbf{T}}^{\perp}}(\overline{\bs{\Psi}})\|_{*} \le \tfrac{k}{4\sqrt{2}} \| \mathcal{P}_{\overline{\mathbf{T}}^{\perp}}(\overline{\bs{\Psi}})\|_{F}, 
\end{align}
in which it can be concluded that as long as $\alpha_3 \leq \tfrac{1}{4}$, $\mathcal{P}_{{\overline{\mathbf T}}^\perp}(\overline{\bs{\Psi}}) = 0$.
Thus, according to the aforementioned arguments, the error bound of $\overline{\bs{\Psi}}$ is given by:
\begin{align}
&\| \overline{\bs{\Psi}} \|_{F} \le  \| \mathcal{P}_{\overline{\mathbf{T}}^{\perp}}(\overline{\bs{\Psi}}) \|_{F} + \| \mathcal{P}_{\overline{\mathbf{T}}}(\overline{\bs{\Psi}}) \|_{F} \nonumber \\
& \le (\tfrac{\sqrt{2}}{k}+1) \| \mathcal{P}_{\overline{\mathbf{T}}^{\perp}}(\overline{\bs{\Psi}}) \|_{F} = 0 ;
\end{align}
and $\overline{\mathbf{ X}} = \mathbf{ X}$. Extending the existing results for low rank matrix and observations in noisy environments, one may write:
\begin{align}
&\| \widehat{\mathbf{X}} - \mathbf{X} \|_{F} \le \tfrac{\sqrt{h}}{k}   \| \mathbf{Q}_{\widetilde{\bm{\mathcal{U}}}_{r}} \mathbf{X}_{r^{+}} \mathbf{Q}_{\widetilde{\bm{\mathcal{V}}}_{r}} \|_{*} + \tfrac{e\sqrt{n}h^{\tfrac{3}{2}}}{k} \nonumber \\
&\le \tfrac{\sqrt{h}}{k} \| \mathbf{Q}_{\widetilde{\bm{\mathcal{U}}}_{r}}\| \| \mathbf{X}_{r^{+}} \|_{*} \| \mathbf{Q}_{\widetilde{\bm{\mathcal{V}}}_{r}}\| + \tfrac{e\sqrt{n}h^{\tfrac{3}{2}}}{k} \nonumber \\
& \le \tfrac{\sqrt{h}}{k} \| \mathbf{X}_{r^{+}} \|_{*} + \tfrac{e\sqrt{n}h^{\tfrac{3}{2}}}{k}.
\end{align} 
\end{proof}  
\section{Proof of Lemma \ref{lemmaA-2-6}}\label{proof lem 3}
As mentioned in Lemma 3, $\mathbf{U}_r$ and $\widetilde{\mathbf{U}}_{r^{\prime}}$ are orthogonal bases of subspaces $\bm{\mathcal{U}}_{r} $ and $ \widetilde{\bm{\mathcal{U}}}_{r^{\prime}} $, respectively. Now without loss of  generality, suppose that  
	\begin{align}
	\mathbf{U}_r^{\rm H} \widetilde{\mathbf{U}}_{r^{\prime}} = [{\rm cos}\bs{\theta}_u \quad \mathbf{0}_{r \times r^{\prime}-r} ]\triangleq \mathbf{U}_r^{\rm H}[\widetilde{\mathbf{U}}_{1,r} \quad \widetilde{\mathbf{U}}_{2,r^{\prime} - r}] \in \mathbb{R}^{r \times r^{\prime}} 
	\end{align}
where $ \widetilde{\mathbf{U}}_{1,r} \in \mathbb{R}^{n \times r} $ and  $ \widetilde{\mathbf{U}}_{2,r^{\prime} - r} \in \mathbb{R}^{ n \times r^{\prime} - r } $ are orthogonal bases for subspaces $ \widetilde{\bm{\mathcal{U}}}_{1,r} \subset \widetilde{\bm{\mathcal{U}}}_{r^{\prime}} $ and $ \widetilde{\bm{\mathcal{U}}}_{2,r^{\prime} -r } \subset \widetilde{\bm{\mathcal{U}}}_{r^{\prime}} $.

To construct orthonormal bases $\mathbf{B}_L$, 
we set  
	\begin{align}
	&\mathbf{U}^{\prime}_{1,r} \triangleq -(\mathbf{I} - \mathbf{U}_r\mathbf{U}_r^{\rm H}) \widetilde{\mathbf{U}}_{1,r}{\rm sin}^{-1}\bs{\theta}_u \nonumber \\
 &=  -\mathbf{P}_{\widetilde{\bm{\mathcal{U}}}^{\perp}_{r}}\widetilde{\mathbf{U}}_{1,r}{\rm sin}^{-1}\bs{\theta}_u \in \mathbb{R}^{n \times r}  \nonumber \\
 	&\mathbf{U}^{\prime}_{2,r^{\prime}-r} \triangleq -(\mathbf{I} - \mathbf{U}_r\mathbf{U}_r^{\rm H}) \widetilde{\mathbf{U}}_{2,r^{\prime}-r} \nonumber \\
  & = -\mathbf{P}_{\widetilde{\bm{\mathcal{U}}}^{\perp}_{r}}\widetilde{\mathbf{U}}_{2,r^{\prime}-r} \in \mathbb{R}^{n \times r^{\prime}-r}
 	\end{align}
 and consider  $$ {\rm span}(\mathbf{U}_{n-r^{\prime} - r}^{\prime \prime}) = {\rm span}([\mathbf{U}_r \quad \mathbf{U}_{r^{\prime}}^{\prime}])^{\perp}. $$
While we presented results and proofs for the column space, it is important to note that all these results are valid for row spaces as well.


\section{Proof of Lemma \ref{lem 4}}\label{proof lemma 4}
We rely on two crucial points for proving the relations in Lemma \ref{lem 4}:
\begin{enumerate}
	\item The operator norm of the diagonal matrix is the maximum element of the matrix. 
	\item For $ \mathbf{X} \in \mathbb{R}^{n \times n }$
	$$ \| \mathbf{X} \| =  \sqrt{\lambda_{\max}(\mathbf{X}^{\rm H}\mathbf{X})} = \sigma_{\max}(\mathbf{X}),$$  
	where $\lambda_{\max}(\mathbf{X}^{\rm H}\mathbf{X})$ is the largest eigenvalue of $\mathbf{X}^{\rm H}\mathbf{X}$ and $\sigma_{\max}(\mathbf{X})$ the largest
	singular value of $\mathbf{X}$ \cite[Lemma A.5.]{foucart2017mathematical}
\end{enumerate}
It then follows that:
	\begin{align}
& \| \mathbf{L}_{11} \| = \| \bs{\Delta}_L\| = \sqrt{\max_{i} \{f^2_2(\lambda_1(i) , \theta_u(i))\}},  \nonumber \\
& \|\mathbf{L}_{12}\| 
=\sqrt{\max_{i}\Big\{ \tfrac{f^2_2(1-\lambda^2_{1}(i) , \theta_u(i))}{f^2_2(\lambda_1(i) , \theta_u(i))}\Big\}},   \nonumber \\
& \|\mathbf{I}_r - \mathbf{L}_{22} \| = \| \mathbf{I}_r - \bs{\Lambda}\mathbf{\Delta}_L^{-1}\| = \nonumber \\
& \sqrt{\max_i\Big\{\bigg(\tfrac{\lambda_{1}(i)-f_2(\lambda_1(i) , \theta_u(i))}{f^2(\lambda_1(i) , \theta_u(i))}\bigg)^2\Big\}},  \nonumber \\
& \Big\|\begin{bmatrix}
\mathbf{I}_{r} - \mathbf{L}_{22} & \\
& \mathbf{I}_{r} - \bs{\Lambda}_{2}
\end{bmatrix} \Big\| =  \max_i \{  \max_i \{1- \lambda_{2}(i)\}\nonumber \\ & \quad \quad \quad \quad \quad \quad , \max_i \Big\{1- \tfrac{\lambda_{1}(i)}{f_2(\lambda_1(i) , \theta_u(i))} \Big\}\}, \nonumber \\
&\|[\mathbf{L}_{11} \quad  \mathbf{L}_{12}]\|^2 = \max_i \Bigg\|\Bigg[
f_2(\lambda_1(i) , \theta_u(i))\nonumber \\
&\quad \quad  \quad \quad \quad \quad  \quad \quad \tfrac{f^2_2\Big((1-\lambda^2_{1}(i)) , \theta_u(i)\Big)}{f_2(\lambda_1(i) , \theta_u(i))}
\Bigg]\Bigg\|^2_{2} \nonumber \\
&= \max_i \big\{ \tfrac{f^2_1(\lambda_1(i) , \theta_u(i))}{f^2_2(\lambda_1(i) , \theta_u(i)) }\big\} \nonumber \\
&\Bigg\|\begin{bmatrix}
\mathbf{0}_r & \mathbf{L}_{12} & &  \\
& \mathbf{L}_{22}-\mathbf{I}_{r} & & \\
& & \bs{\Lambda}_2 - \mathbf{I}_{r^{\prime}-r} & \\
& & &\mathbf{0}_{n-r^{\prime}-r}
\end{bmatrix}\Bigg\|^{2} = \nonumber \\
&\max \Big\{ \Bigg\| \begin{bmatrix}
\mathbf{L}_{12} \\  \mathbf{L}_{22}-\mathbf{I}_{r}
\end{bmatrix}\Bigg\|_2^2  , \| \bs{\Lambda}_2 - \mathbf{I}_{r^{\prime}-r} \|_2^2 \Big\}	
\nonumber \\
& =\max_i \{[ \max_i \Big\{1- (\tfrac{\lambda_{1}(i)}{f_2(\lambda_1(i) , \theta_u(i))})^2 + \nonumber \\
&\quad \quad \tfrac{f^2_2(1-\lambda^2_{1}(i) , \theta_u(i))}{f^2_2(\lambda_1(i) , \theta_u(i))} \Big\} , \max_i \{(\lambda_{2}(i)-1)^2\}] \}.
	\end{align}
	\end{appendices}
\section*{Acknowledgement}
The author H. F. Ardakani would like to thank N. Rahmani for proofreading an initial version of this draft.
	\bibliographystyle{ieeetr}
	\bibliography{MyrefrenceMRadMC}
\end{document}